\pdfoutput=1

\documentclass[12pt,a4paper]{article}

\usepackage{ifthen} 
\newboolean{pdflatex}
\setboolean{pdflatex}{true} 

\newboolean{articletitles}
\setboolean{articletitles}{true} 

\newboolean{uprightparticles}
\setboolean{uprightparticles}{false} 

\newboolean{inbibliography}
\setboolean{inbibliography}{false} 

\usepackage{microtype}
\usepackage{lineno}  
\usepackage{xspace} 
\usepackage{caption} 

\usepackage{graphicx}  
\usepackage{color}
\usepackage{colortbl}
\graphicspath{{./figs/}} 

\usepackage{amsmath} 
\usepackage{amssymb}
\usepackage{amsfonts}
\usepackage{upgreek} 

\newcommand*\patchAmsMathEnvironmentForLineno[1]{%
\expandafter\let\csname old#1\expandafter\endcsname\csname #1\endcsname
\expandafter\let\csname oldend#1\expandafter\endcsname\csname
end#1\endcsname
 \renewenvironment{#1}%
   {\linenomath\csname old#1\endcsname}%
   {\csname oldend#1\endcsname\endlinenomath}%
}
\newcommand*\patchBothAmsMathEnvironmentsForLineno[1]{%
  \patchAmsMathEnvironmentForLineno{#1}%
  \patchAmsMathEnvironmentForLineno{#1*}%
}
\AtBeginDocument{%
\patchBothAmsMathEnvironmentsForLineno{equation}%
\patchBothAmsMathEnvironmentsForLineno{align}%
\patchBothAmsMathEnvironmentsForLineno{flalign}%
\patchBothAmsMathEnvironmentsForLineno{alignat}%
\patchBothAmsMathEnvironmentsForLineno{gather}%
\patchBothAmsMathEnvironmentsForLineno{multline}%
}

\usepackage{hyperref}    
\usepackage[all]{hypcap} 

\usepackage{placeins}
\usepackage{overpic}

\def\lhcb {\mbox{LHCb}\xspace}

\def\babar  {\mbox{BaBar}\xspace}

\def\MagUp {\mbox{\em Mag\kern -0.05em Up}\xspace}

\ifthenelse{\boolean{uprightparticles}}%
{
 \def\Pbeta       {\ensuremath{\upbeta}\xspace}
                  
 \def\Pdelta      {\ensuremath{\updelta}\xspace}

 \def\Ppi         {\ensuremath{\uppi}\xspace}

 \def\PDelta      {\ensuremath{\Delta}\xspace}                 
 \def\PXi      {\ensuremath{\Xi}\xspace}                 
 \def\PLambda      {\ensuremath{\Lambda}\xspace}                 
 \def\PSigma      {\ensuremath{\Sigma}\xspace}                 
 \def\POmega      {\ensuremath{\Omega}\xspace}                 
 \def\PUpsilon      {\ensuremath{\Upsilon}\xspace}                 
 
 \def\PB      {\ensuremath{\mathrm{B}}\xspace}                 
                  
 \def\PD      {\ensuremath{\mathrm{D}}\xspace}

 \def\PK      {\ensuremath{\mathrm{K}}\xspace}

 \def\Pb      {\ensuremath{\mathrm{b}}\xspace}                 
 \def\Pc      {\ensuremath{\mathrm{c}}\xspace}

 \def\Ph      {\ensuremath{\mathrm{h}}\xspace}                 
 \def\Pi      {\ensuremath{\mathrm{i}}\xspace}

 \def\Ps      {\ensuremath{\mathrm{s}}\xspace}

}
{
 \def\Pbeta       {\ensuremath{\beta}\xspace}
                  
 \def\Pdelta      {\ensuremath{\delta}\xspace}

 \def\Ppi         {\ensuremath{\pi}\xspace}

 \mathchardef\PDelta="7101
 \mathchardef\PXi="7104
 \mathchardef\PLambda="7103
 \mathchardef\PSigma="7106
 \mathchardef\POmega="710A
 \mathchardef\PUpsilon="7107
                  
 \def\PB      {\ensuremath{B}\xspace}                 
                  
 \def\PD      {\ensuremath{D}\xspace}

 \def\PK      {\ensuremath{K}\xspace}

 \def\Pb      {\ensuremath{b}\xspace}                 
 \def\Pc      {\ensuremath{c}\xspace}

 \def\Ph      {\ensuremath{h}\xspace}                 
 \def\Pi      {\ensuremath{i}\xspace}

 \def\Ps      {\ensuremath{s}\xspace}

}

\makeatletter
\ifcase \@ptsize \relax
  \newcommand{\miniscule}{\@setfontsize\miniscule{4}{5}}
\or
  \newcommand{\miniscule}{\@setfontsize\miniscule{5}{6}}
\or
  \newcommand{\miniscule}{\@setfontsize\miniscule{5}{6}}
\fi
\makeatother

\DeclareRobustCommand{\optbar}[1]{\shortstack{{\miniscule (\rule[.5ex]{1.25em}{.18mm})}
  \\ [-.7ex] $#1$}}

\def\squark    {{\ensuremath{\Ps}}\xspace}

\def\cquark    {{\ensuremath{\Pc}}\xspace}

\def\bquark    {{\ensuremath{\Pb}}\xspace}

\def\pion   {{\ensuremath{\Ppi}}\xspace}

\def\pip    {{\ensuremath{\pion^+}}\xspace}
\def\pim    {{\ensuremath{\pion^-}}\xspace}
\def\pipm   {{\ensuremath{\pion^\pm}}\xspace}

\def\kaon    {{\ensuremath{\PK}}\xspace}
  \def\Kbar    {{\kern 0.2em\overline{\kern -0.2em \PK}{}}\xspace}

\def\KorKbar    {\kern 0.18em\optbar{\kern -0.18em K}{}\xspace}
\def\Kz      {{\ensuremath{\kaon^0}}\xspace}
\def\Kzb     {{\ensuremath{\Kbar{}^0}}\xspace}
\def\Kp      {{\ensuremath{\kaon^+}}\xspace}
\def\Km      {{\ensuremath{\kaon^-}}\xspace}
\def\Kpm     {{\ensuremath{\kaon^\pm}}\xspace}

\def\KS      {{\ensuremath{\kaon^0_{\rm\scriptscriptstyle S}}}\xspace}
\def\KL      {{\ensuremath{\kaon^0_{\rm\scriptscriptstyle L}}}\xspace}

  \def\Dbar    {{\kern 0.2em\overline{\kern -0.2em \PD}{}}\xspace}
\def\D       {{\ensuremath{\PD}}\xspace}

\def\DorDbar    {\kern 0.18em\optbar{\kern -0.18em D}{}\xspace}
\def\Dz      {{\ensuremath{\D^0}}\xspace}
\def\Dzb     {{\ensuremath{\Dbar{}^0}}\xspace}

\def\Dstar   {{\ensuremath{\D^*}}\xspace}

\def\Dstarpm {{\ensuremath{\D^{*\pm}}}\xspace}

\def\B       {{\ensuremath{\PB}}\xspace}
\def\Bbar    {{\ensuremath{\kern 0.18em\overline{\kern -0.18em \PB}{}}}\xspace}

\def\BorBbar    {\kern 0.18em\optbar{\kern -0.18em B}{}\xspace}

\def\Bu      {{\ensuremath{\B^+}}\xspace}
\def\Bub     {{\ensuremath{\B^-}}\xspace}
\def\Bp      {{\ensuremath{\Bu}}\xspace}
\def\Bm      {{\ensuremath{\Bub}}\xspace}
\def\Bpm     {{\ensuremath{\B^\pm}}\xspace}

\def\Bd      {{\ensuremath{\B^0}}\xspace}
\def\Bs      {{\ensuremath{\B^0_\squark}}\xspace}

  \def\Y#1S{\ensuremath{\PUpsilon{(#1S)}}\xspace}

\def\Lbar        {{\ensuremath{\kern 0.1em\overline{\kern -0.1em\PLambda}}}\xspace}
\def\LorLbar    {\kern 0.18em\optbar{\kern -0.18em \PLambda}{}\xspace}

\def\to                 {\ensuremath{\rightarrow}\xspace}

\def\CP                {{\ensuremath{C\!P}}\xspace}

\def\AT#1     {\ensuremath{A_{\mathrm{T}}^{#1}}\xspace}           

\def\C#1      {\ensuremath{\mathcal{C}_{#1}}\xspace}                       
\def\Cp#1     {\ensuremath{\mathcal{C}_{#1}^{'}}\xspace}                    
\def\Ceff#1   {\ensuremath{\mathcal{C}_{#1}^{\mathrm{(eff)}}}\xspace}        
\def\Cpeff#1  {\ensuremath{\mathcal{C}_{#1}^{'\mathrm{(eff)}}}\xspace}       
\def\Ope#1    {\ensuremath{\mathcal{O}_{#1}}\xspace}                       
\def\Opep#1   {\ensuremath{\mathcal{O}_{#1}^{'}}\xspace}                    

\newcommand{\tev}{\ifthenelse{\boolean{inbibliography}}{\ensuremath{~T\kern -0.05em eV}\xspace}{\ensuremath{\mathrm{\,Te\kern -0.1em V}}}\xspace}
\newcommand{\gev}{\ensuremath{\mathrm{\,Ge\kern -0.1em V}}\xspace}
\newcommand{\mev}{\ensuremath{\mathrm{\,Me\kern -0.1em V}}\xspace}
\newcommand{\kev}{\ensuremath{\mathrm{\,ke\kern -0.1em V}}\xspace}
\newcommand{\ev}{\ensuremath{\mathrm{\,e\kern -0.1em V}}\xspace}
\newcommand{\gevc}{\ensuremath{{\mathrm{\,Ge\kern -0.1em V\!/}c}}\xspace}
\newcommand{\mevc}{\ensuremath{{\mathrm{\,Me\kern -0.1em V\!/}c}}\xspace}
\newcommand{\gevcc}{\ensuremath{{\mathrm{\,Ge\kern -0.1em V\!/}c^2}}\xspace}
\newcommand{\gevgevcccc}{\ensuremath{{\mathrm{\,Ge\kern -0.1em V^2\!/}c^4}}\xspace}
\newcommand{\mevcc}{\ensuremath{{\mathrm{\,Me\kern -0.1em V\!/}c^2}}\xspace}

\def\mum  {\ensuremath{{\,\upmu\rm m}}\xspace}

\def\invfb   {\ensuremath{\mbox{\,fb}^{-1}}\xspace}

\newcommand{\chisq}{\ensuremath{\chi^2}\xspace}
\newcommand{\chisqndf}{\ensuremath{\chi^2/\mathrm{ndf}}\xspace}
\newcommand{\chisqip}{\ensuremath{\chi^2_{\rm IP}}\xspace}

\def\gsim{{~\raise.15em\hbox{$>$}\kern-.85em
          \lower.35em\hbox{$\sim$}~}\xspace}
\def\lsim{{~\raise.15em\hbox{$<$}\kern-.85em
          \lower.35em\hbox{$\sim$}~}\xspace}

\def\sPlot{\mbox{\em sPlot}\xspace}

\def\pt         {\mbox{$p_{\rm T}$}\xspace}

\def\evtgen     {\mbox{\textsc{EvtGen}}\xspace}

\def\geant      {\mbox{\textsc{Geant4}}\xspace}

\def\photos     {\mbox{\textsc{Photos}}\xspace}

\def\pythia     {\mbox{\textsc{Pythia}}\xspace}

\def\tell1  {TELL1\xspace}
\def\ukl1   {UKL1\xspace}

\newcommand{\BtoDpi}{\ensuremath{\Bpm\to\PD\pipm}\xspace}
\newcommand{\BtoDK}{\ensuremath{\Bpm\to\PD\Kpm}\xspace}

\newcommand{\Dpi}{\ensuremath{\PD\pi^\pm}\xspace}

\newcommand{\DK}{\ensuremath{\PD\PK^\pm}\xspace}

\DeclareRobustCommand{\titleoptbar}[1]{\shortstack{{\miniscule (\rule[.5ex]{1.75em}{.18mm})}   \\ [-.3ex] $#1$}}
\DeclareRobustCommand{\titleoptbarlowcase}[1]{\shortstack{{\miniscule (\rule[.5ex]{1.25em}{.18mm})}   \\ [-.3ex] $#1$}}
\DeclareRobustCommand{\optbarlowcase}[1]{\shortstack{{\miniscule (\rule[.5ex]{1.0em}{.18mm})}   \\ [-.7ex] $#1$}}
\def\nuornubar{\optbarlowcase{\nu}{} \xspace}

\newcommand{\BpmtoDpipm}{\ensuremath{\Bpm\to\PD\pipm}\xspace}
\newcommand{\BpmtoDKpm}{\ensuremath{\Bpm\to\PD\Kpm}\xspace}
\newcommand{\BpmtoDhpm}{\ensuremath{\Bpm\to\PD\Ph^\pm}\xspace}

\newcommand{\TitleBztoDstmu}{\ensuremath{\titleoptbar{B}{} \to\Dstarpm\mu^\mp\titleoptbarlowcase{\nu}\!_\mu}\xspace}
\newcommand{\BztoDstmu}{\ensuremath{\BorBbar\!\to\Dstarpm\mu^\mp\nuornubar_{\!\!\mu}}\xspace}

\newcommand{\BztoDstmuX}{\ensuremath{\BorBbar\!\to\Dstarpm\mu^\mp\nuornubar_{\!\!\mu} X}\xspace}

\newcommand{\KsPiPi}{\ensuremath{\KS\pip\pim}\xspace}
\newcommand{\KsKK}{\ensuremath{\KS\Kp\Km}\xspace}
\newcommand{\Kshh}{\ensuremath{\KS\Ph^+\Ph^-}\xspace}

\newcommand{\DtoKsPiPi}{\ensuremath{\PD\to\KsPiPi}\xspace}
\newcommand{\DtoKsKK}{\ensuremath{\PD\to\KsKK}\xspace}
\newcommand{\DtoKshh}{\ensuremath{\PD\to\Kshh}\xspace}

\newcommand{\BpmtoDhpmDtoKspipi}{\ensuremath{\Bpm\to\PD(\KsPiPi)\Ph^\pm}\xspace}

\newcommand{\BpmtoDhpmDtoKsKK}{\ensuremath{\Bpm\to\PD(\KsKK)\Ph^\pm}\xspace}

\newcommand{\etaDP}{\ensuremath{\varepsilon}\xspace}
\newcommand{\etaDst}{\ensuremath{\varepsilon_{\Dstar\!\mu}}\xspace}
\newcommand{\etaDK}{\ensuremath{\varepsilon_{DK}}\xspace}

\newcommand{\xpm}{\ensuremath{x_{\pm}}\xspace}
\newcommand{\ypm}{\ensuremath{y_{\pm}}\xspace}
\def\xy {\ensuremath{x_{\pm}},\xspace \ensuremath{y_{\pm}}\xspace}

\usepackage{cite} 
\usepackage{mciteplus}

\usepackage{longtable} 

\begin{document}

\renewcommand{\thefootnote}{\fnsymbol{footnote}}
\setcounter{footnote}{1}


\begin{titlepage}
\pagenumbering{roman}

\vspace*{-1.5cm}
\centerline{\large EUROPEAN ORGANIZATION FOR NUCLEAR RESEARCH (CERN)}
\vspace*{1.5cm}
\hspace*{-0.5cm}
\begin{tabular*}{\linewidth}{lc@{\extracolsep{\fill}}r}
\ifthenelse{\boolean{pdflatex}}
{\vspace*{-2.7cm}\mbox{\!\!\!\includegraphics[width=.14\textwidth]{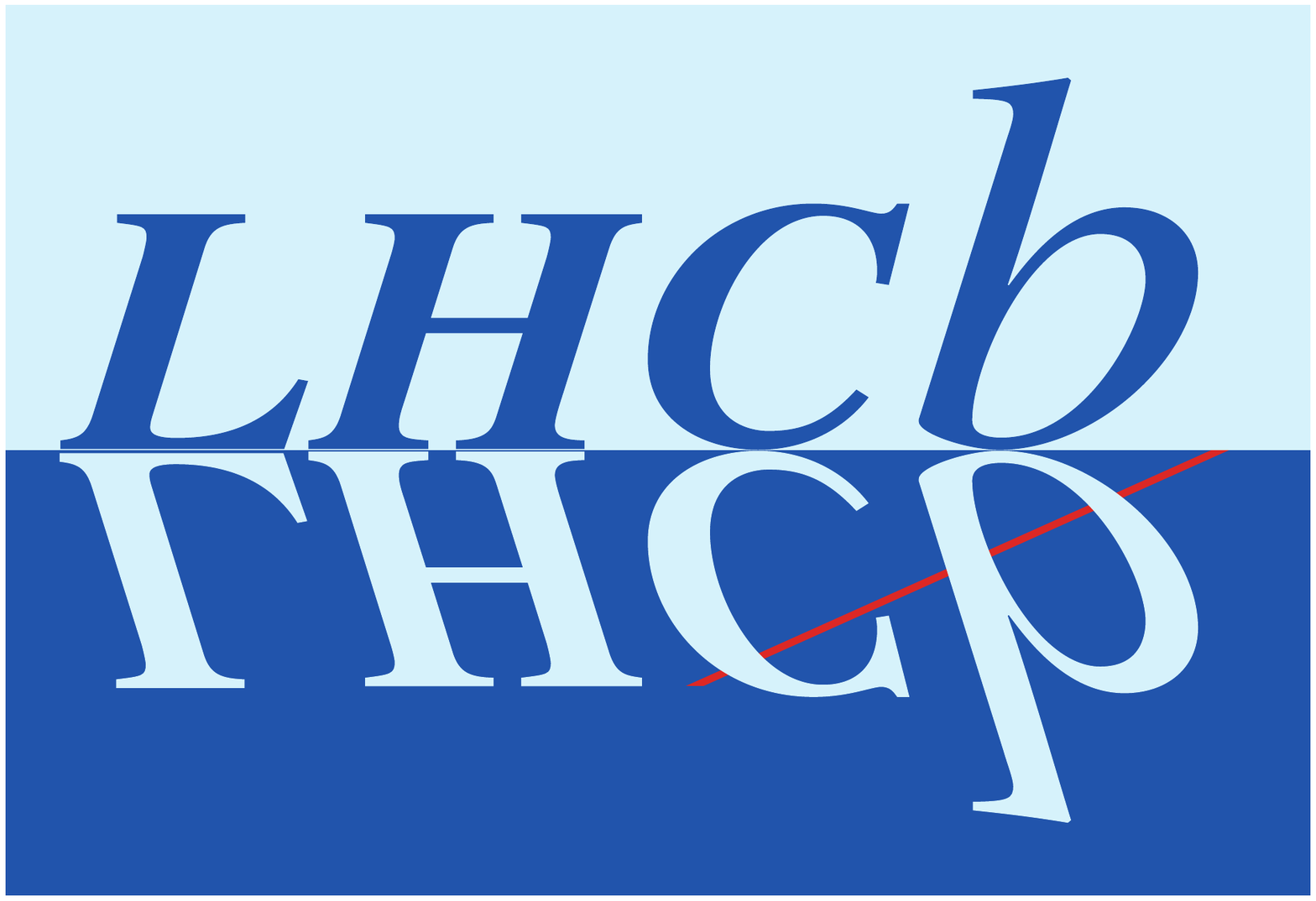}} & &}%
{\vspace*{-1.2cm}\mbox{\!\!\!\includegraphics[width=.12\textwidth]{lhcb-logo.eps}} & &}%
\\
 & & CERN-PH-EP-2014-202 \\  
 & & LHCb-PAPER-2014-041 \\  
 & & 12 August 2014\\ 
 & & \\
\end{tabular*}

\vspace*{4.0cm}

{\bf\boldmath\huge
\begin{center}
Measurement of the CKM angle $\gamma$ \\ 
using \BtoDK with \\ 
$D \to \KsPiPi, \KsKK$ decays
\end{center}
}


\begin{center}
The LHCb collaboration\footnote{Authors are listed at the end of this paper.}
\end{center}


\begin{abstract}
  \noindent
 A binned Dalitz plot analysis of $B^\pm \to D K^\pm$ decays, with $D \to \KS \pi^+\pi^-$ and $D \to \KS  K^+ K^-$, is performed to measure the \CP-violating observables $x_{\pm}$ and $y_{\pm}$, which are sensitive to the Cabibbo-Kobayashi-Maskawa angle $\gamma$.  The analysis exploits a sample of proton-proton collision data corresponding to 3.0\invfb collected by the LHCb experiment. Measurements from CLEO-c of the variation of the strong-interaction phase of the $D$ decay over the Dalitz plot are used as inputs. The values of the parameters are found to be $x_+   =   ( -7.7 \pm 2.4 \pm 1.0 \pm 0.4 )\times 10^{-2}$, $x_- =   (2.5 \pm 2.5 \pm 1.0 \pm 0.5) \times 10^{-2}$,   $y_+  = (-2.2 \pm 2.5 \pm 0.4 \pm 1.0)\times 10^{-2}$, and $y_-  = (7.5 \pm 2.9 \pm 0.5 \pm 1.4) \times 10^{-2}$. The first, second, and third uncertainties are the statistical, the experimental systematic, and that associated with the precision of the strong-phase parameters. These are the most precise measurements of these observables and correspond to $\gamma = (62^{\,+15}_{\,-14})^\circ$, 
 with a second solution at $\gamma \to \gamma + 180^\circ$, 
and $r_B = 0.080^{+ 0.019}_{-0.021}$, where $r_B$ is the ratio between the suppressed and favoured $B$ decay amplitudes.
\end{abstract}


\begin{center}
Published in JHEP 10 (2014) 097
\end{center}

\vspace{\fill}

{\footnotesize 
\centerline{\copyright~CERN on behalf of the \lhcb collaboration, license \href{http://creativecommons.org/licenses/by/4.0/}{CC-BY-4.0}.}}
\vspace*{2mm}

\end{titlepage}


\newpage
\setcounter{page}{2}
\mbox{~}
%

\cleardoublepage


\renewcommand{\thefootnote}{\arabic{footnote}}
\setcounter{footnote}{0}



\pagestyle{plain} 
\setcounter{page}{1}
\pagenumbering{arabic}

\section{Introduction}
A precise determination of the Cabibbo-Kobayashi-Maskawa (CKM) angle $\gamma \equiv \arg(-V_{\rm ud} V^*{}_{\rm \!\!\!\!\! ub} / V_{\rm cd} V^*{}_{\rm \!\!\!\!\! cb})$  
is of great value in testing the Standard Model (SM) description of \CP violation. Measurements of this weak phase in tree-level processes involving the interference between  $b \to c \bar{u}s$ and $b \to u \bar{c} s$ transitions are expected to be insensitive to contributions from physics beyond the SM. Such measurements therefore provide a SM benchmark against which other observables, more likely to be affected by physics beyond the SM, can be compared.  The effects of the interference can be probed by studying \CP-violating observables in 
\BpmtoDKpm decays, where $D$ represents a neutral $D$ meson reconstructed in a final state that is common to both \Dz and \Dzb decays.  Examples of such final states recently studied by LHCb are two-body decays~\cite{LHCb-PAPER-2012-001}, multibody decays that are not self-conjugate~\cite{LHCb-PAPER-2012-055,LHCb-PAPER-2013-068}, and self-conjugate three-body decays, such as \KsPiPi and \KsKK, designated collectively as \Kshh~\cite{LHCbGGSZ1fb}. Similar measurements have also been made using neutral \Bd~\cite{LHCb-PAPER-2014-028} and \Bs~\cite{LHCb-PAPER-2014-038} mesons. 

Sensitivity to $\gamma$ in \BtoDK, \DtoKshh decays is obtained by comparing the distribution of the events in the \DtoKshh Dalitz plot for reconstructed $B^+$ and $B^-$ mesons~\cite{GGSZ,BONDARGGSZ}.  
Knowledge of the variation of the strong-interaction phase of the $D$ decay over the Dalitz plot is required to determine $\gamma$. 
One approach, adopted by the \babar~\cite{BABAR2005,BABAR2008,BABAR2010}, Belle~\cite{BELLE2004,BELLE2006,BELLE2010}  and LHCb~\cite{LHCb-PAPER-2014-017} collaborations, is to use an amplitude model determined from flavour-tagged \DtoKshh decays to provide this input.
An attractive alternative~\cite{GGSZ,BPMODIND1,BPMODIND2} is to use direct measurements of the strong-phase variation over bins of the Dalitz plot, thereby avoiding 
model-related systematic uncertainties. 
Such measurements can be obtained using quantum-correlated $\Dz\Dzb$ pairs from $\psi(3770)$ decays and have been made at CLEO-c~\cite{CLEOCISI}.  This model-independent method has been applied to measurements at Belle~\cite{BELLEMODIND} using \BpmtoDKpm, \DtoKsPiPi decays, and at LHCb~\cite{LHCbGGSZ1fb} using a subset of data used in the current analysis. 

In this paper, $pp$ collision data at a centre-of-mass energy $\sqrt{s}=7\ (8)$~\tev, accumulated by \lhcb in 2011 (2012) and corresponding to a total integrated luminosity of $3.0\invfb$, are exploited to perform a model-independent study of the decay mode \BpmtoDKpm with \DtoKsPiPi and \DtoKsKK. In addition to benefiting from a larger data set than that used in Ref.~\cite{LHCbGGSZ1fb} the current study makes use of improved analysis techniques. The results presented here thus supersede those of Ref.~\cite{LHCbGGSZ1fb}.

\section{Overview of the analysis}
\label{sec:formalism}

The amplitude of the decay $B^- \to D K^-$, \DtoKshh can be written as a superposition of the $B^- \to D^0 K^-$ and  $B^- \to \Dzb K^-$ contributions, given by  
\begin{equation}
A_B (m_-^2, m_+^2) \propto \, A + r_B e^{i(\delta_B - \gamma)}\overline{A}.
\label{eq:bamplitude}
\end{equation}
Here  $m_-^2$ and $m_+^2$ are the  invariant masses squared of the $\KS h^-$ and  $\KS h^+$ combinations, respectively,
that define the position of the decay in the Dalitz plot, $A=A(m_-^2,m_+^2)$ is the $\Dz \to \KS h^+ h^-$ amplitude 
and $\overline{A}=\overline{A}(m_-^2,m_+^2)$ the  $\Dzb \to \KS h^+ h^-$ amplitude. 
The parameters $r_B$ and $\delta_B$ are the ratio of the magnitudes of the $B^- \to \Dzb K^-$ and $B^- \to \Dz K^-$ amplitudes, and the strong-phase difference between them.  
The equivalent expression for the charge-conjugated decay $B^+ \to D K^+$ is obtained by making the substitutions $\gamma \to -\gamma$ and $A \leftrightarrow \overline{A}$.  Neglecting \CP violation in charm decays, which is known to be small in \Dz--\Dzb mixing and Cabibbo-favoured $D$ meson decays~\cite{PDG2012}, the conjugate amplitudes are related by $A(m_-^2,m_+^2) = \overline{A}(m_+^2,m_-^2)$.

The Dalitz plot is partitioned into $2N$ regions symmetric under the exchange $m_+^2 \leftrightarrow m_-^2$, following Ref.~\cite{GGSZ}.  The bins are labelled from $-N$ to $+N$ (excluding zero), where the positive bins have $m_-^2 > m_+^2$.  At each point in the Dalitz plot, there is a strong-phase difference $\delta_D(m_-^2,m_+^2) \equiv \arg A - \arg \overline{A}$ between the \Dz and $\Dzb$ decay.  The cosine of the strong-phase difference averaged in each bin and weighted by the decay rate is termed $c_i$ and is given by
\begin{equation}
c_i \equiv \frac{\int_{{\cal D}_i} (|A||\overline{A}| \cos{\delta_D})\, d{\cal D}}
{\sqrt{\int_{{\cal D}_i} |A|^2 \,d{\cal D}} \,\sqrt{ \int_{{\cal D}_i} |\overline{A}|^2 \,d{\cal D}}},
\end{equation}
where the integrals are evaluated over the area ${\cal D}_i$ of bin $i$.  An analogous expression may be written for $s_i$, which is the 
sine of the strong-phase difference within bin $i$, weighted by the decay rate.  
The values of $c_i$ and $s_i$ have been directly measured by the CLEO collaboration, exploiting quantum-correlated $\Dz\Dzb$ pairs produced at the $\psi(3770)$ resonance~\cite{CLEOCISI}. 
One $D$ meson was reconstructed in a decay to either $\KS h^+ h^-$ or $\KL h^+ h^-$, and the other $D$ meson was reconstructed either in a \CP eigenstate or in a decay to \Kshh. The efficiency-corrected event yields, combined with flavour-tag information, allowed $c_i$ and $s_i$ to be determined. 
There is a systematic uncertanty associated with using these direct measurements due their finite precision. The alternative is to calculate $c_i$ and $s_i$ assuming a functional form for $|A|$, $|\overline{A}|$ and $\delta_D$, which may be obtained from an amplitude model fitted to flavour-tagged $D^0$ decays. This alternative method relies on assumptions about the nature of the intermediate resonances that contribute to the \Kshh final state, and leads to a systematic uncertainty associated with the variation in $\delta_D$.  

\begin{figure}[t]
\begin{center}
\includegraphics[width=0.48\textwidth]{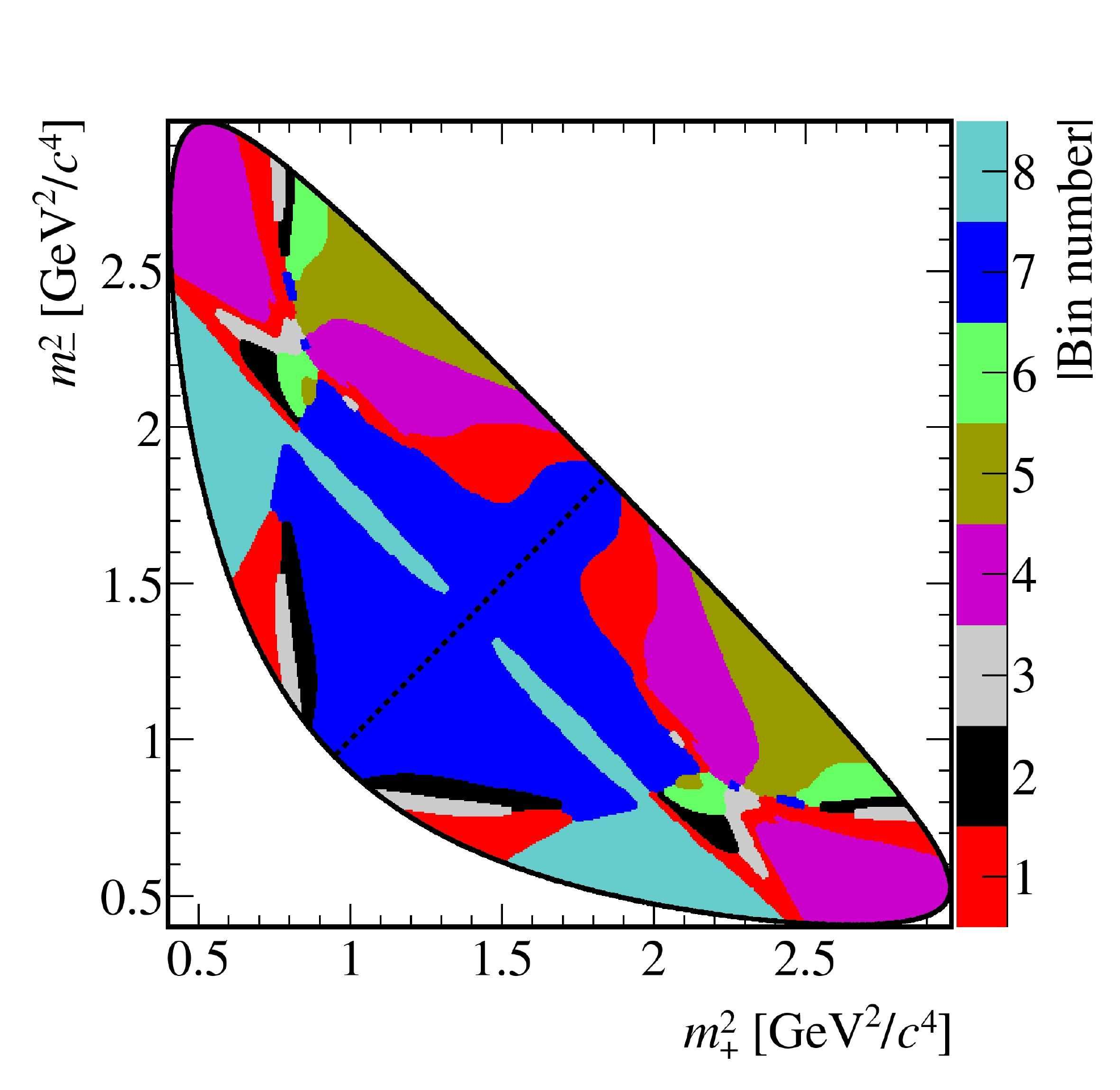}
\includegraphics[width=0.48\textwidth]{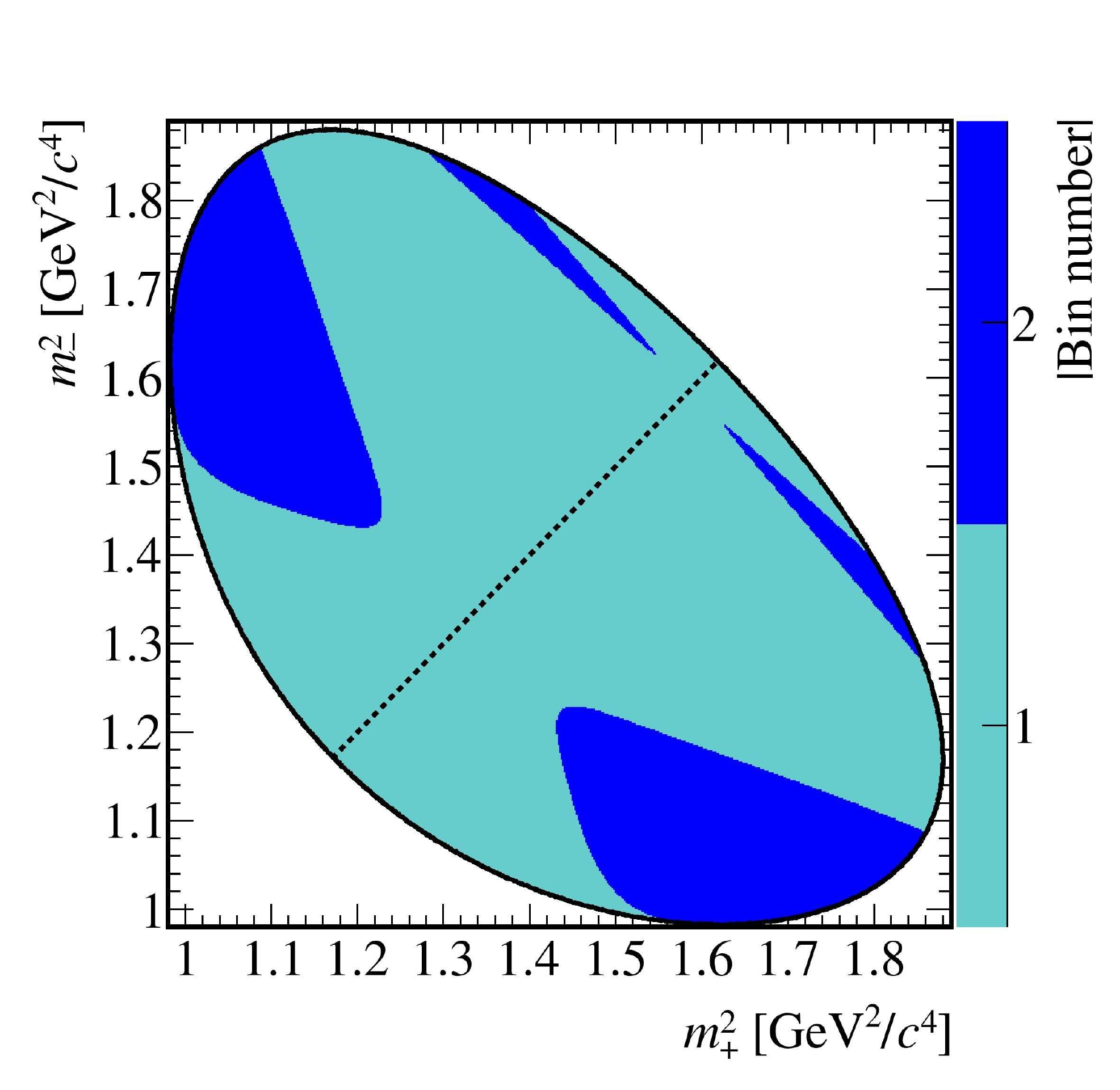}
\caption{\small Binning schemes for (left) \DtoKsPiPi and (right) \DtoKsKK. The diagonal line separates the positive and negative bins, where the positive bins are in the region where $m^2_->m^2_+$ is satisfied. }
\label{fig:bins}
\end{center}
\end{figure}

In the CLEO-c study the \KsPiPi Dalitz plot was partitioned into $2 \times 8$ bins, with a number of schemes available. The `optimal binning' variant~\cite{CLEOCISI}, where the bins have been chosen to optimise the statistical sensitivity to $\gamma$, is adopted in this analysis. The optimisation was performed assuming a strong-phase difference distribution as given by the \babar model presented in Ref.~\cite{BABAR2008}. For the \KsKK final state, $c_i$ and $s_i$ measurements are available for the Dalitz plot partitioned into different numbers of bins with the guiding model being that from the \babar study described in Ref.~\cite{BABAR2010}. 
The analysis described here adopts the $2 \times 2$ option, a decision driven by the size of the signal sample. 
The use of a specific model in defining the bin boundaries does not bias the $c_i$ and $s_i$ measurements.  If the model is a poor description of the underlying decay the only consequence is a reduction in the statistical sensitivity of the $\gamma$ measurement. 
The binning choices for the two decay modes are shown in Fig.~\ref{fig:bins}.

The population of each positive (negative) bin in the Dalitz plot arising from $B^+$ decays is $N_{+i}^+$ ($N_{-i}^+$), and that from $B^-$ decays is $N_{+i}^-$ ($N_{-i}^-$). The physics parameters of interest, $r_B$, $\delta_B$, and $\gamma$, are translated into four \CP observables~\cite{BABAR2005} that are measured in this analysis. These observables are defined as
\begin{equation}
x_\pm \equiv r_B \cos (\delta_B \pm \gamma) {\rm\ \ and\ } \; y_\pm \equiv r_B \sin (\delta_B \pm \gamma).
\label{eq:xydefinitions}
\end{equation}

The selection requirements introduce nonuniformities in the populations of the Dalitz plot.
The relative selection and reconstruction efficiency profile $\varepsilon=\varepsilon(m^2_-,m^2_+)$ for signal candidates is defined as a function of the position in the Dalitz plot. 
The absolute normalisation of \etaDP is not relevant; only the efficiency associated with one point relative to the others matters. 
Considering Eq.~\ref{eq:bamplitude} it follows that 
\begin{eqnarray}
N_{\pm i}^+ &=& h_{B^+} \left[ F_{\mp i} + (x_+^2 + y_+^2) F_{\pm i} + 2 \sqrt{F_i F_{-i}} ( x_+ c_{\pm i} - y_+ s_{\pm i}) \right], \nonumber \\
N_{\pm i}^- &=& h_{B^-} \left[ F_{\pm i} + (x_-^2 + y_-^2) F_{\mp i} + 2 \sqrt{F_i F_{-i}} ( x_- c_{\pm i} + y_- s_{\pm i}) \right],
\label{eq:populations2}
\end{eqnarray}
where the value $F_i$ is given by
\begin{equation}
\label{eq:fi}
F_i = \frac{\int_{{\cal D}_i} |A|^2\,\etaDP \, d{\cal D}}{\sum_j \int_{{\cal D}_j}|A|^2\,\etaDP \, d{\cal D}}
\end{equation}
and is the fraction of events in bin $i$ of the $\Dz\to\Kshh$ Dalitz plot. 
The quantities $h_{B^\pm}$ are normalisation factors, which can be different for $B^+$ and $B^-$ due to asymmetries in production rates of bottom and antibottom mesons.

The observed distribution of candidates over the \DtoKshh Dalitz plot is used to fit for $x_\pm$, $y_\pm$ and $h_{B^\pm}$. The values of $F_i$ are determined from the control mode \BztoDstmuX, where the \Dstarpm decays to $\DorDbar^0\pipm$, and the $\DorDbar^0$ decays to either the \KsPiPi or \KsKK final state. The symbol $X$, hereinafter omitted, indicates other particles that are potentially produced in the \BorBbar decay. Samples of simulated events are used to correct for differences in the efficiency for reconstructing and selecting \BztoDstmu and \BtoDK decays.

In addition to selecting \BpmtoDKpm and \BztoDstmu candidates we also select \BpmtoDpipm decays. Candidates selected in this decay mode provide an important control sample that is used to constrain the invariant mass shape of the \BpmtoDKpm signal and to determine the yield of \BpmtoDpipm decays misidentified as \BpmtoDKpm candidates. 

The use of \BztoDstmu decays to determine the values of $F_i$ is an improvement over Ref.~\cite{LHCbGGSZ1fb}, for which the decay \BpmtoDpipm was used. 
The small level of \CP violation in the latter decay led to a significant systematic uncertainty. 
This uncertainty is eliminated when using the flavour-specific semileptonic decay. There is still a systematic uncertainty associated with the procedure but it is relatively small in magnitude. 

The effect of \Dz--\Dzb mixing is ignored in the above discussion, and was neglected in the CLEO-c measurements of $c_i$ and $s_i$ as well as in the values of $F_i$. This leads to a bias of approximately $0.2^\circ$ in the $\gamma$ determination~\cite{BPV}, which is negligible for the current analysis. The effect of \CP violation in \KS decays is expected to lead to a $\mathcal{O}(1^\circ)$ uncertainty~\cite{Yuval}, and is also ignored given the expected precision. An uncertainty due to the different nuclear interaction cross sections for \Kz and \Kzb mesons is expected to be of a similar magnitude and is also ignored~\cite{LHCb-PAPER-2014-013}.

The rest of the paper is organised as follows. Section~\ref{sec:Detector} describes the LHCb detector, and Section~\ref{sec:selection} presents the selection and the model used to fit the invariant mass spectrum. Sections~\ref{sec:dmu} and~\ref{sec:corr} are concerned with the selection of the semileptonic control channel, used to determine the signal efficiency profile.  Section~\ref{sec:analysis} discusses the binned Dalitz plot fit and presents the results for the \CP parameters.  The evaluation of systematic uncertainties is summarised in Section~\ref{sec:syst}.  In Section~\ref{sec:discussion} the use of the measured \CP parameters to determine the CKM angle $\gamma$ is described. The results of the analysis are summarised in Section~\ref{sec:conclusions}.

\section{Detector and simulation}
\label{sec:Detector}

The \lhcb detector~\cite{Alves:2008zz} is a single-arm forward
spectrometer covering the \mbox{pseudorapidity} range $2<\eta <5$,
designed for the study of particles containing \bquark or \cquark
quarks. The detector includes a high-precision tracking system
consisting of a silicon-strip vertex detector surrounding the $pp$
interaction region, a large-area silicon-strip detector located
upstream of a dipole magnet with a bending power of about
$4{\rm\,Tm}$, and three stations of silicon-strip detectors and straw
drift tubes~\cite{LHCb-DP-2013-003} placed downstream.
The combined tracking system provides a momentum measurement with
relative uncertainty that varies from 0.4\% at 2\gevc to 0.6\% at 100\gevc,
and impact parameter resolution of 20\mum for
tracks with large transverse momentum. Different types of charged hadrons are distinguished using information
from two ring-imaging Cherenkov detectors~\cite{LHCb-DP-2012-003}. Photon, electron and
hadron candidates are identified by a calorimeter system consisting of
scintillating-pad and preshower detectors, an electromagnetic
calorimeter and a hadronic calorimeter. Muons are identified by a
system composed of alternating layers of iron and multiwire
proportional chambers~\cite{LHCb-DP-2012-002}.
The trigger~\cite{LHCb-DP-2012-004} consists of a
hardware stage, based on information from the calorimeter and muon
systems, followed by a software stage, which applies a full event
reconstruction. 
The trigger algorithms used to select candidate fully hadronic and semileptonic \PB decays 
are slightly different due to the presence of the muon in the latter. 

In the simulation, $pp$ collisions are generated using
\pythia~\cite{Sjostrand:2006za,Sjostrand:2007gs} 
 with a specific \lhcb
configuration~\cite{LHCb-PROC-2010-056}.  Decays of hadronic particles
are described by \evtgen~\cite{Lange:2001uf}, in which final-state
radiation is generated using \photos~\cite{Golonka:2005pn}. The
interaction of the generated particles with the detector and its
response are implemented using the \geant
toolkit~\cite{Allison:2006ve, Agostinelli:2002hh} as described in
Ref.~\cite{LHCb-PROC-2011-006}.

\section{Event selection and fit to invariant mass spectrum for
\texorpdfstring{$\boldsymbol{\BpmtoDKpm}$}{B to DK} and \texorpdfstring{$\boldsymbol{\BpmtoDpipm}$ decays}{B to Dpi}
}
\label{sec:selection}

Selection requirements are applied to obtain an event sample enriched with \BpmtoDKpm and \BpmtoDpipm candidates, where $D$ indicates a \Dz or \Dzb meson that decays to the final state \Kshh. 
The kaon or pion produced directly in the \Bpm decay is denoted the `bachelor' hadron. 
Decays of \KS mesons to the $\pip\pim$ final state are reconstructed in two different categories,
the first involving \KS mesons that decay early enough for the pions to be reconstructed in the vertex detector, the
second containing \KS that decay later such that track segments of the pions cannot be formed in the vertex detector. These categories are
referred to as \emph{long} and \emph{downstream}, respectively. The candidates in the long category have better mass, momentum and vertex resolution 
than those in the downstream category. Henceforth \Bpm candidates are denoted long or downstream depending on which \KS type they contain.

Events considered in the analysis must fulfil both hardware and software trigger requirements. 
At the hardware stage at least one of the two following criteria must be satisfied: either a particle produced in the decay of the signal \Bpm candidate leaves a deposit with high transverse energy in the hadronic calorimeter, 
or the event is accepted because particles not associated with the signal candidate fulfil the trigger requirements. 
The software trigger designed to select \BpmtoDKpm and \BpmtoDpipm candidates requires a two-, three- or four-track secondary vertex with a large sum of the transverse momentum, \pt, of the associated charged particles and a significant displacement from the primary $pp$ interaction vertices~(PVs). At least one charged particle should have $\pt > 1.7\gevc$ and \chisqip with respect to any primary interaction greater than 16, where \chisqip is defined as the   difference in \chisq of a given PV fitted with and without the considered track. 
A multivariate algorithm~\cite{BBDT} is used for the identification of secondary vertices that are consistent with the decay of a \bquark hadron.

A multivariate approach is employed to improve the event selection relative to that used in Ref.~\cite{LHCbGGSZ1fb}. A boosted decision tree~\cite{Breiman, AdaBoost} (BDT) 
is trained on simulated signal events and background taken from the high \Bpm mass sideband (5800--7000\mevcc). 
Both signal and background samples contain candidates from the \PD and \KS signal regions only. 
Different BDTs are trained for long and downstream candidates. 
Each BDT uses the following variables: the logarithm of the \chisqip of the pions from the \D decay and also of the bachelor particle; the logarithm of the \chisqip of the \KS decay products (long candidates only); the logarithm of the \D \chisqip; the \Bpm \chisqip; a variable characterising the \Bpm flight distance; 
the \Bpm and \D momenta; the \chisq of the kinematic fit of the whole decay chain, (described in detail below); and the `\Bpm isolation variable', a quantity designed to ensure the \Bpm candidate is well isolated from other tracks in the event. The \Bpm isolation variable is the asymmetry between the \pt of the signal candidate and the vector sum of the \pt of the other tracks in the event that lie within a distance of 1.5 rad in $\eta$--$\phi$ space, where $\phi$ is the azimuthal angle. 
The discriminating power of the variables differs slightly for long and downstream candidates. Two variables that are highly discriminating for both samples are the \Bpm \chisqip and \Bpm isolation variable.
An optimal value of the BDT discriminator is determined with a series of pseudo-experiments to obtain the value that provides the best sensitivity to \xy. Events in the data sample that have a value below the optimum are rejected. The optimal BDT value is different for long and downstream candidates primarily because the level of combinatorial background is larger for the latter.

To suppress background further, the \KS, \D and \Bpm momentum vectors are required to point in the same direction as the vector connecting their production and decay vertices. 
The mass of the \D candidate must lie within 25\mevcc of the known \D mass~\cite{PDG2012}.

Particle identification (PID) requirements are placed on the bachelor to separate \BpmtoDKpm and \BpmtoDpipm candidates. PID criteria are also applied to the kaons from the \D decay for the final state \KsKK. To ensure good control of the PID performance it is required that information from the RICH detectors is present.

A kinematic fit~\cite{Hulsbergen:2005pu} is imposed on the full \Bpm decay chain. The fit constrains the \Bpm candidate to point towards the PV and the \D and \KS candidates to have their known masses~\cite{PDG2012}. 
This fit improves the \Bpm mass resolution and therefore provides greater discrimination between signal and background; furthermore, it improves the resolution on the Dalitz plot and ensures that all candidates lie within the kinematically-allowed region of the Dalitz plot.  
The candidates obtained in this fit are used to determine the physics parameters of interest. 
An additional fit, in which only the \Bpm pointing and \D mass constraints are imposed, is employed to aid discrimination between genuine and background \KS candidates. After this fit is applied it is required that the mass of the \KS candidate lies within 15\mevcc of its known value~\cite{PDG2012}. 

To remove background from $D\to\pip\pim\pip\pim$ decays, long \KS candidates are required to have travelled a significant distance from the \D vertex. To remove charmless \Bpm decays, the displacement along the beamline between the \D and \Bpm decay vertices is required to be positive.

\begin{figure}[t]
\centering
\includegraphics[width=0.98\textwidth]{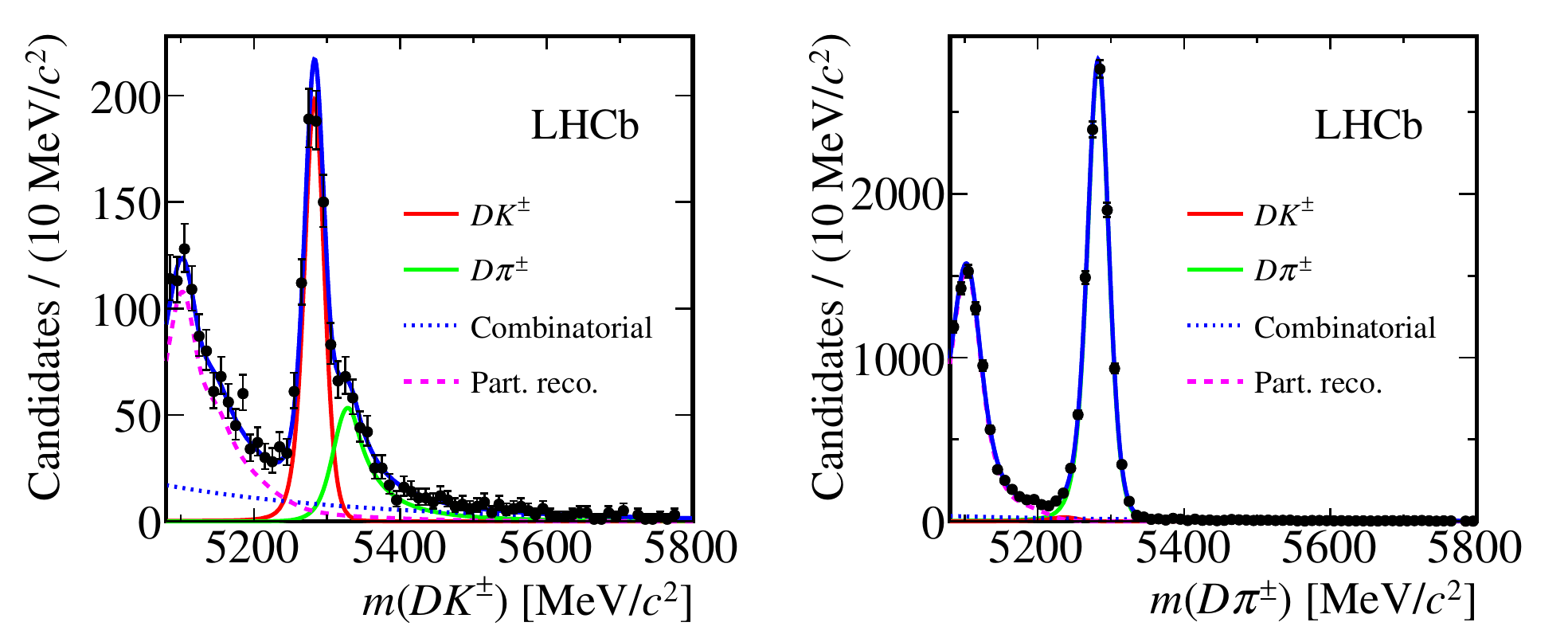} \\
\includegraphics[width=0.98\textwidth]{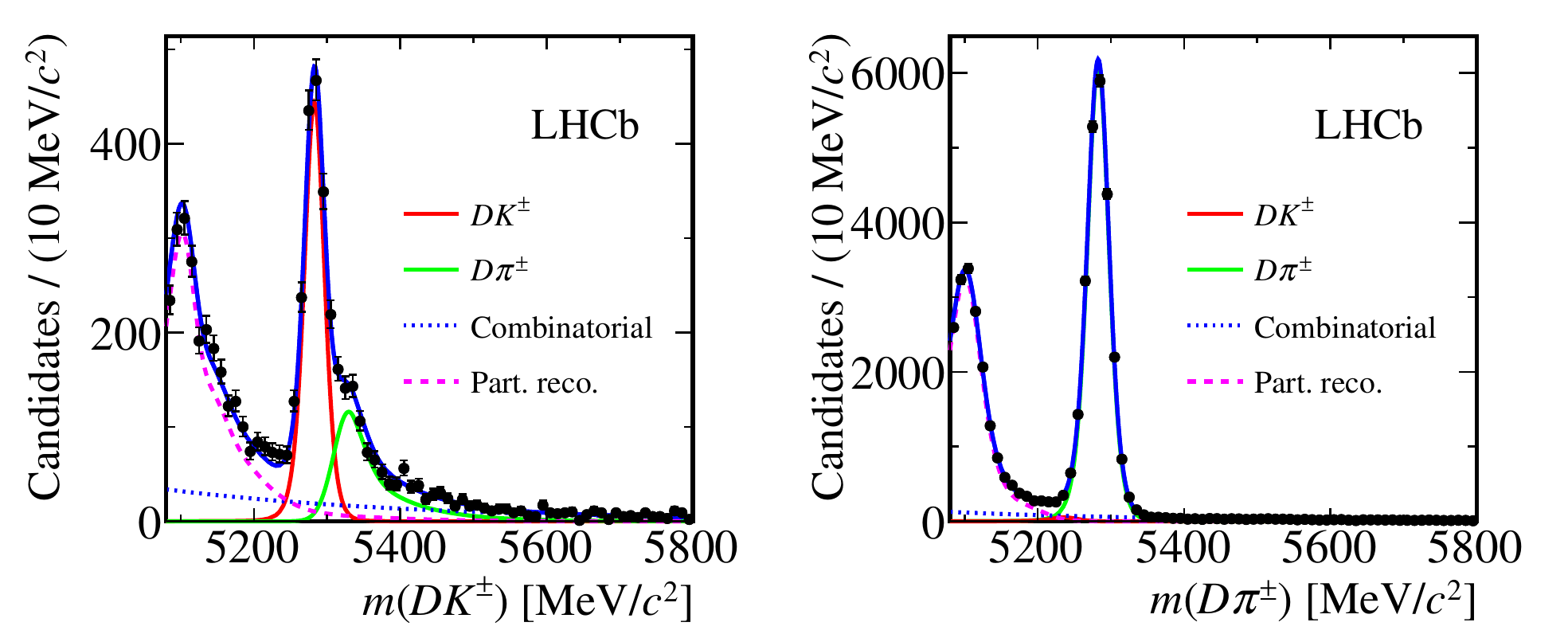}
\caption{\small Invariant mass distributions of (left) \BpmtoDKpm and  (right) \BpmtoDpipm candidates, with \DtoKsPiPi, divided between the (top) long  and (bottom) downstream \KS categories. Fit results, including the signal and background components, are superimposed.}
\label{fig:mass_kspipi}
\end{figure}

\begin{figure}[t]
\centering
\includegraphics[width=0.98\textwidth]{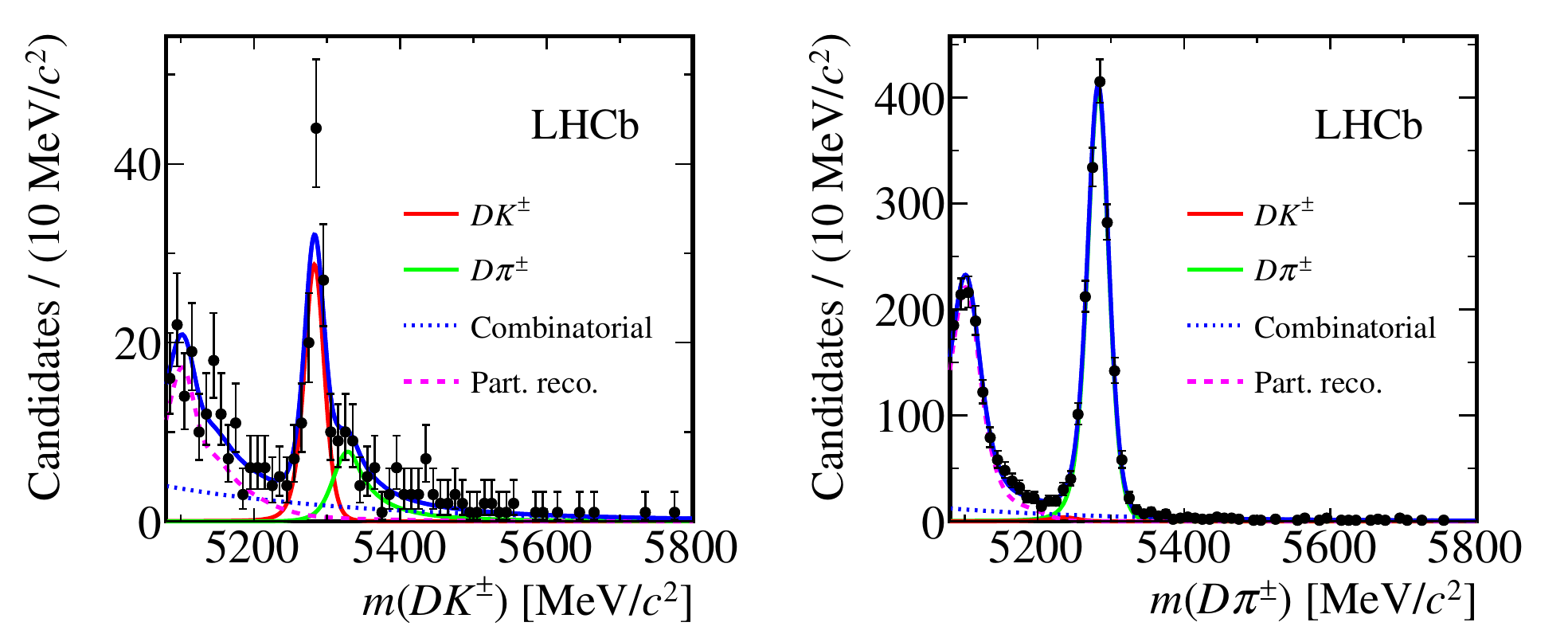} \\
\includegraphics[width=0.98\textwidth]{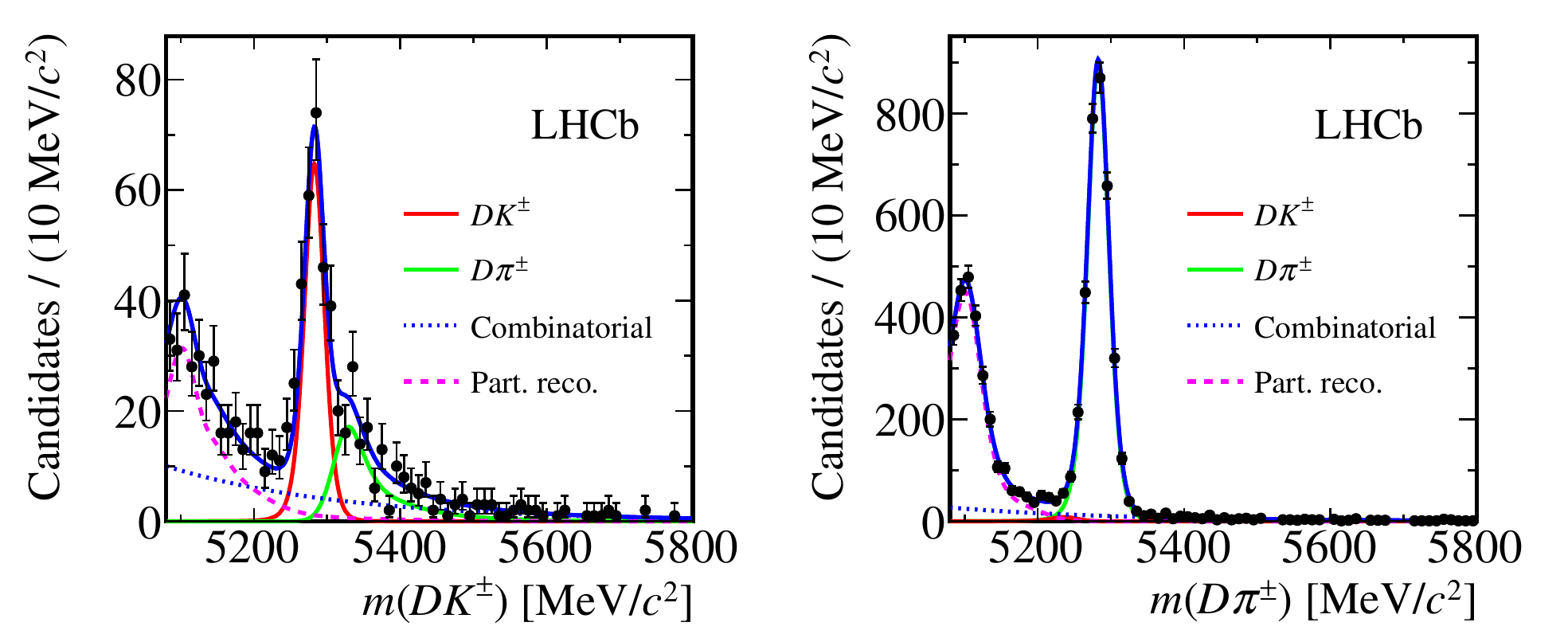}
\caption{\small Invariant mass distributions of (left) \BpmtoDKpm and  (right) \BpmtoDpipm candidates, with \DtoKsKK, divided between the (top) long  and (bottom) downstream \KS categories. Fit results, including the signal and background components, are superimposed.}
\label{fig:mass_kskk}
\end{figure}

The invariant mass distributions of the selected candidates are shown in Fig.~\ref{fig:mass_kspipi} for \BpmtoDKpm and \BpmtoDpipm, with \DtoKsPiPi decays, divided between the long and downstream \KS categories. Figure~\ref{fig:mass_kskk} shows the corresponding distributions for final states with \DtoKsKK. 
The result of an extended maximum likelihood fit to these distributions is superimposed.   
The fit is performed simultaneously for  \BpmtoDKpm and \BpmtoDpipm candidates, including both \DtoKsPiPi and 
\DtoKsKK decays, allowing several independent parameters for long and downstream \KS categories.   
The fit range is between 5080\mevcc and 5800\mevcc in the \Bpm invariant mass. 
The purpose of this simultaneous fit to data integrated over the Dalitz plot is to determine the parameters that describe the invariant mass spectrum in preparation for the binned fit described in Sect.~\ref{sec:analysis}.  
The mass spectrum of \BtoDpi candidates is fitted because it is similar to the \BtoDK spectrum, aiding the determination of the signal lineshape due to the higher yield and lower background. The yield of \BtoDpi candidates misidentified as \BtoDK candidates can be determined from knowledge of the \BtoDpi signal yield and the PID selection efficiencies. 

The signal probability density function (PDF) is a Gaussian function with asymmetric tails, defined as
\begin{equation}
f(m; m_0,\sigma,\alpha_L,\alpha_R) \propto 
  \begin{cases}
   \exp[-(m-m_0)^2/(2\sigma^2 + \alpha_L(m-m_0)^2)] & \  m < m_0 \\
   \exp[-(m-m_0)^2/(2\sigma^2 + \alpha_R(m-m_0)^2)] & \  m > m_0
  \end{cases}
\label{eq:cruijff}
\end{equation}
where $m$ is the candidate mass and $m_0$, $\sigma$, $\alpha_L$, and $\alpha_R$ are free parameters in the fit. The parameter $m_0$ is common to all classes of signal. The parameters describing the asymmetric tails, $\alpha_{L,R}$, are fitted separately for events with long and downstream \KS categories. The width parameter $\sigma$ is left as a free parameter for the two \KS categories, but the ratio between this width in \BpmtoDKpm and \BpmtoDpipm decays is required to be the same, independent of \KS reconstruction or \Dz decay category. 
The width is determined to be around 13\mevcc  for \BpmtoDKpm decays of both \KS classes,  and is 10\% larger for \BpmtoDpipm decays. The yield of \BpmtoDpipm candidates in each category is determined in the fit. Instead of fitting the yield of the \BpmtoDKpm candidates separately, the ratio $\mathcal{R} \equiv N(\BpmtoDKpm)$/$N(\BpmtoDpipm)$ is determined and is constrained to have the same value for all categories. 

The background has contributions from random track combinations and partially reconstructed \PB decays.  The random track combinations are modelled by exponential PDFs. The slopes of these functions are determined through the study of two independent samples: candidates reconstructed such that both charged hadrons produced in the \PD decay have the same sign, and candidates reconstructed using the \PD mass sidebands. 
The slopes are consistent with each other. In the fit to the signal data the exponential slopes are Gaussian-constrained to the results of the sideband studies. 

A significant background component exists in the \BpmtoDKpm sample, 
arising from a fraction of the dominant \BpmtoDpipm decays in which the bachelor particle is misidentified as a kaon by the RICH system.  
The yield of this type of background is calculated using knowledge of misidentification efficiencies that are obtained from large samples of kinematically selected $\Dstarpm \to \DorDbar^0 \pi^\pm$, $\DorDbar^0 \to K^\mp \pi^\pm$ decays.   The tracks in this calibration sample are reweighted to match the momentum and pseudorapidity distributions of the bachelor tracks in the \Bpm decay sample, thereby ensuring that the measured PID performance is representative of that in the \Bpm decay sample.  The efficiency to identify a kaon correctly  is found to be 86\%, and that for a pion to be 96\%.  The efficiency of misidentifying a pion as a kaon is 4\%. 
From this information and from the knowledge of the number of  reconstructed \BpmtoDpipm decays, the amount of this background surviving the \BpmtoDKpm selection is estimated.

The distribution of true \BpmtoDpipm candidates misidentified as \BpmtoDKpm candidates is determined using data. The \Bpm invariant mass distribution is obtained by reconstructing candidates in the \BpmtoDpipm sample with a kaon mass hypothesis for the bachelor pion. The sample is weighted using the \sPlot{} method~\cite{Pivk:2004ty} and the PID efficiencies. The use of the \sPlot{} method in the reweighting suppresses partially reconstructed and combinatorial backgrounds. Weighting by PID efficiencies allows for reproduction of the kinematic properties of pions misidentified as kaons in the signal \BpmtoDKpm sample. The weighted distribution is fitted to a parametric shape with different shapes used for the samples containing long and downstream \KS decays.  The fitted parameters are subsequently fixed in the fit to the \Bpm invariant mass spectrum.  

A similar procedure is used to determine the number of \BpmtoDKpm decays misidentified as \BpmtoDpipm. Due to the reduced branching fraction of \BpmtoDKpm and the small likelihood of misidentifying a kaon as a pion, such cases occur at a low rate and have a minor influence on the fit.

Partially reconstructed $b$-hadron decays (shown as Part. Reco. in Fig.~\ref{fig:mass_kspipi} and Fig.~\ref{fig:mass_kskk}) contaminate the sample predominantly at invariant masses smaller than that of the signal peak. 
These decays contain an unreconstructed  pion or a photon, which comes from a vector-meson decay. The dominant decays in the signal region are $B^\pm \to \D \rho^\pm$, $B^\pm \to D^{*0} \pi^\pm$ and $B^0 \to D^{*\pm} \pi^\mp$ decays in which one particle is missed.
The distribution in the invariant mass spectrum depends on the spin and mass of the missing particle. If the missing particle has spin-parity $J^P=0^-$ ($1^-$), the distribution is parameterised with a parabola with positive (negative) curvature convolved with a resolution function. 
The mass of the missing particle defines the kinematic endpoints of the distribution prior to reconstruction. The shapes for decays in which a particle is missed and a pion is misidentified as a kaon are parameterised with semi-empirical PDFs formed from sums of Gaussian and error functions. The parameters of these distributions are fixed to the results of fits to data from two-body $D$ decays, with the exception of the resolution function width, the ratio of widths in the \BpmtoDKpm and \BpmtoDpipm channels and a shift along the \Bpm mass. 
The resulting PDF is cross-checked with a similar fit to an admixture of simulated backgrounds.

The number of \BpmtoDKpm candidates in each \KS category or \Dz decay category is determined from the value of $\mathcal{R}$ and the number of \BpmtoDpipm events in the corresponding category. The ratio $\mathcal{R}$ is determined in the fit and measured to be $(7.7 \pm 0.2)\%$ (statistical uncertainty only), consistent with that observed in Ref.~\cite{LHCb-PAPER-2012-001}. The yields returned by the invariant mass fit in the full fit region are scaled to the signal region, defined as 5247--5317\mevcc, and are presented in Tables~\ref{tab:sigyieldpipi} and~\ref{tab:sigyieldkk}. Because the \BtoDK yields are calculated using $\mathcal{R}$ their uncertainties are smaller than those that would be expected if the yields were allowed to vary in the fit. 

\begin{table}[t]
\centering
\caption{Yields and statistical uncertainties in the signal region from the invariant mass fits, scaled from the full fit mass range, for candidates passing the \BpmtoDhpmDtoKspipi selection. Values are shown separately for candidates formed using long and downstream \KS decays.
The signal region is between 5247\mevcc and 5317\mevcc and the full fit range is between 5080\mevcc and 5800\mevcc. \label{tab:sigyieldpipi}}
\begin{tabular}{l r@{$\ \pm \ $}l r@{$\ \pm \ $}l l r@{$\ \pm \ $}l r@{$\ \pm \ $}l}
\hline
Fit component           & \multicolumn{4}{c}{\BpmtoDKpm selection} & & \multicolumn{4}{c}{\BpmtoDpipm selection} \\
\cline{2-5} \cline{7-10}
                        & \multicolumn{2}{c}{Long} & \multicolumn{2}{c}{Downstream} & & \multicolumn{2}{c}{Long} & \multicolumn{2}{c}{Downstream} \\
\hline
\BpmtoDKpm              & $702$ & $18$ & $1555$ & $39$ & & $   30$ & $  5$ & $   64$ & $7$ \\
\BpmtoDpipm             & $ 87$ & $ 9$ & $ 164$ & $13$ & & $10\,338$ & $106$ & $22\,779$ & $166$ \\
Combinatorial           & $ 59$ & $ 9$ & $ 133$ & $14$ & & $  103$ & $ 11$ & $  433$ & $25$ \\
Partially reconstructed & $ 38$ & $ 2$ & $  82$ & $ 3$ & & $  4.6$ & $0.1$ & $ 14.2$ & $0.1$ \\
\hline
\end{tabular}
\end{table}

\begin{table}[t]
\centering
\caption{Yields and statistical uncertainties in the signal region from the invariant mass fits, scaled from the full fit mass range, for candidates passing the \BpmtoDhpmDtoKsKK selection. Values are shown separately for candidates formed using long and downstream \KS decays.
The signal region is between 5247\mevcc and 5317\mevcc and the full fit range is between 5080\mevcc and 5800\mevcc. \label{tab:sigyieldkk}}
\begin{tabular}{l r@{$\ \pm \ $}l r@{$\ \pm \ $}l l r@{$\ \pm \ $}l r@{$\ \pm \ $}l}
\hline
Fit component           & \multicolumn{4}{c}{\BpmtoDKpm selection} & & \multicolumn{4}{c}{\BpmtoDpipm selection} \\
\cline{2-5} \cline{7-10}
                        & \multicolumn{2}{c}{Long} & \multicolumn{2}{c}{Downstream} & & \multicolumn{2}{c}{Long} & \multicolumn{2}{c}{Downstream} \\
\hline
\BpmtoDKpm              & $101$ & $  4$ & $223$ & $  7$ & & $ 4.5$ & $1.9$  & $10.1$ & $2.9$ \\
\BpmtoDpipm             & $ 13$ & $  3$ & $ 24$ & $  5$ & & $1501$ & $38$   & $3338$ & $57$  \\
Combinatorial           & $ 13$ & $  3$ & $ 30$ & $  5$ & & $  36$ & $ 5$   & $  78$ & $ 7$  \\
Partially reconstructed & $4.6$ & $0.7$ & $8.6$ & $1.2$ & & $0.60$ & $0.02$ & $ 2.0$ & $0.1$ \\
\hline
\end{tabular}
\end{table}

\begin{figure}[t]
\centering
\includegraphics[width=0.45\textwidth]{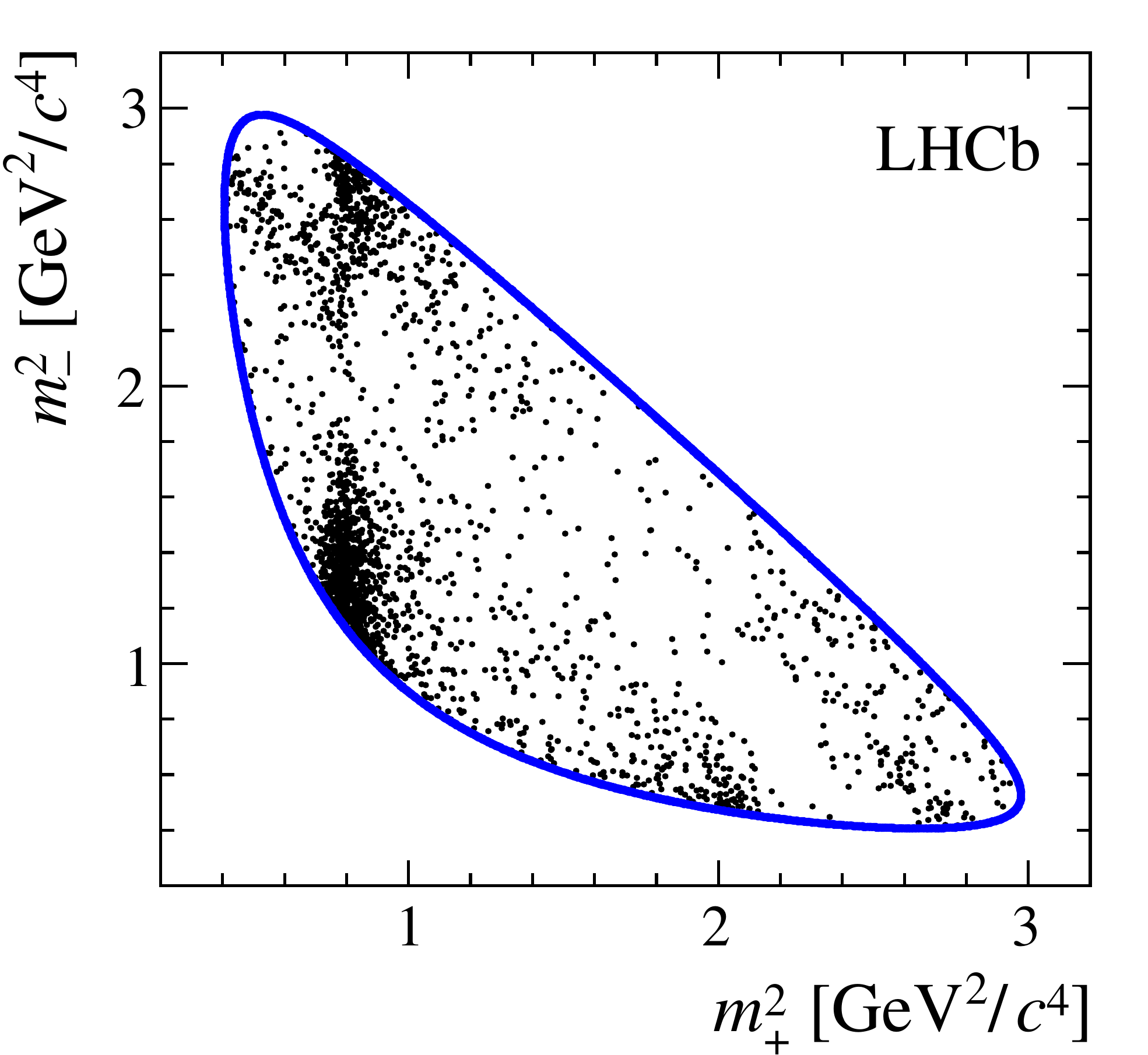}
\includegraphics[width=0.45\textwidth]{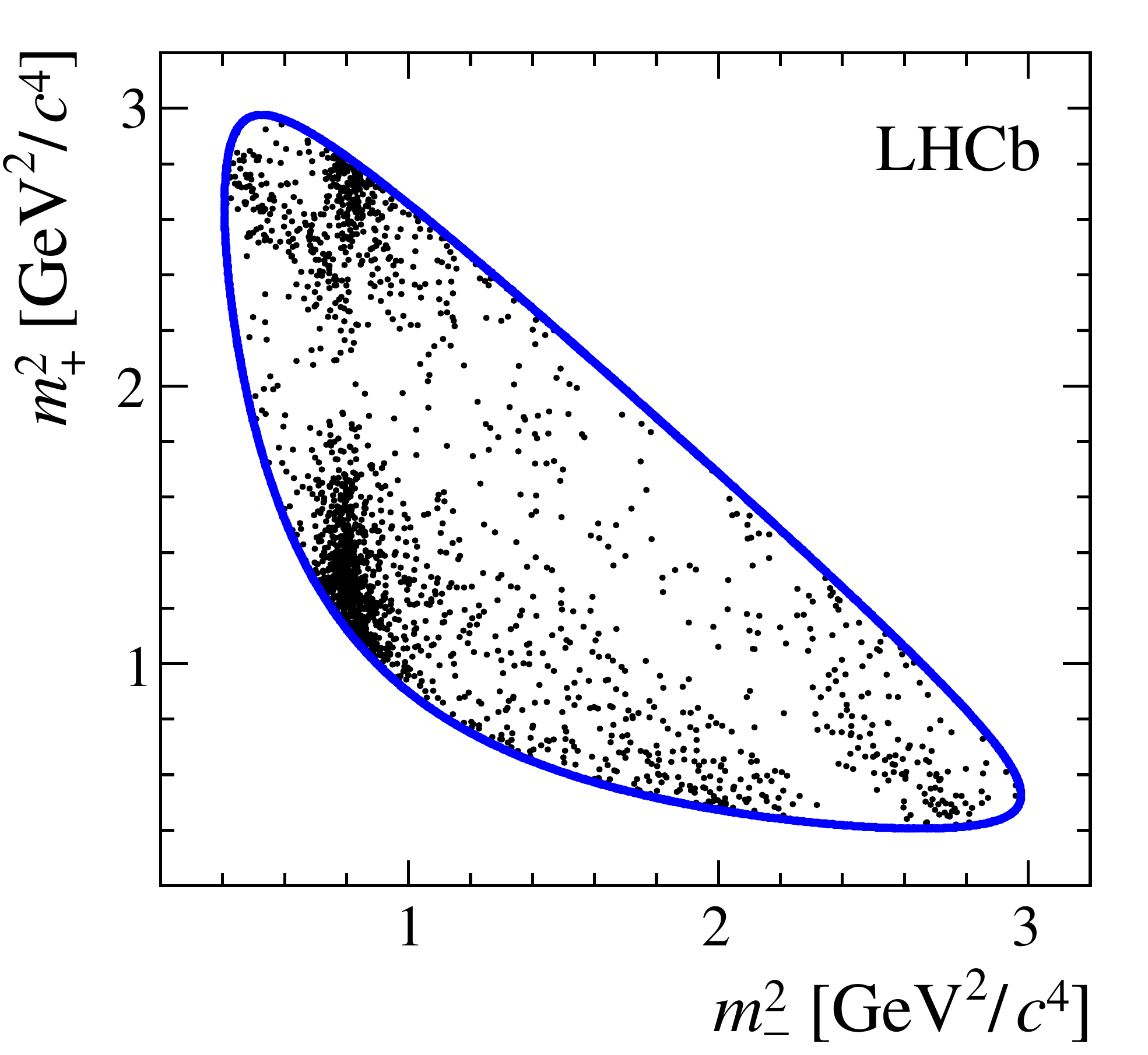}
\caption{\small Dalitz plots of \BpmtoDKpm candidates in the signal region for \DtoKsPiPi decays from (left) \Bp and (right) \Bm decays. 
Both long and downstream \KS candidates are included. The blue line indicates the kinematic boundary.}
\label{fig:dalitzpipi}
\end{figure}

\begin{figure}[t]
\centering
\includegraphics[width=0.45\textwidth]{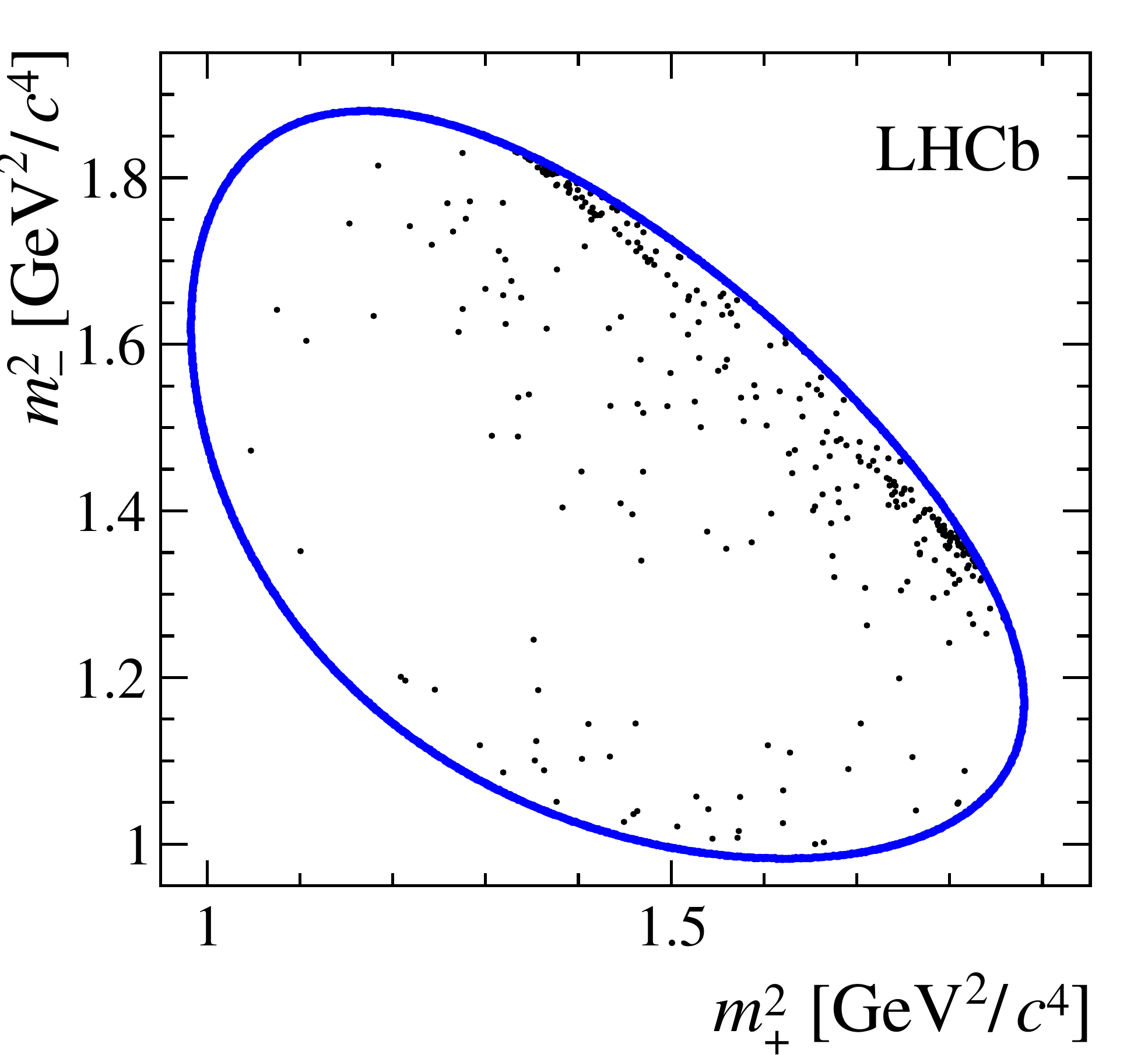}
\includegraphics[width=0.45\textwidth]{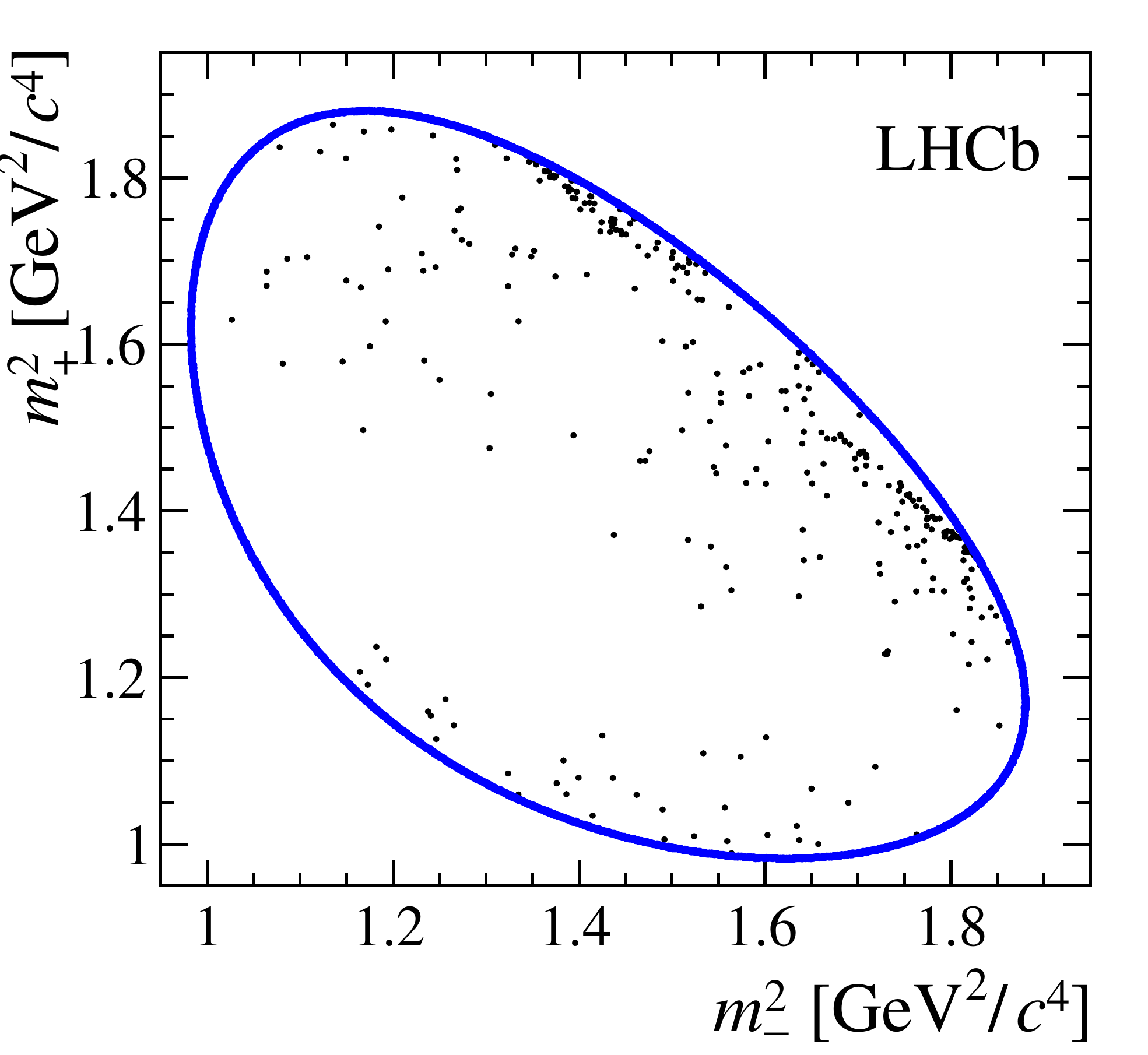}
\caption{\small Dalitz plots of \BpmtoDKpm candidates in the signal region for \DtoKsKK decays from (left) \Bp and (right) \Bm decays. 
Both long and downstream \KS candidates are included. The blue line indicates the kinematic boundary.}
\label{fig:dalitzkk}
\end{figure}

The Dalitz plots for \BpmtoDKpm candidates restricted to the signal region for the two \DtoKshh final states are shown in Figs.~\ref{fig:dalitzpipi} and~\ref{fig:dalitzkk}.  Separate plots are shown for $B^+$ and $B^-$ decays.

\section{Event selection and yield determination for \texorpdfstring{\boldmath${\TitleBztoDstmu}$ decays}{B to Dstar mu}} 

\label{sec:dmu}

A sample of the decays \BztoDstmu, $\Dstarpm \to \DorDbar^0\pipm$, $\DorDbar^0\to\Kshh$ is used to determine the quantities $F_i$. These are defined in Eq.~\ref{eq:fi} as the expected fractions of \Dz decays falling into the Dalitz plot bin labelled $i$, taking into account the efficiency profile of the signal decay.
The semileptonic decay of the $B$ and the strong-interaction decay of the $\Dstarpm$ allow the flavour of the \Dz meson to be determined from the charge of the bachelor muon and pion. This particular decay chain, involving a flavour-tagged \Dz decay, is chosen due to its low background level and low mistag probability. The selection requirements are chosen to minimise changes to the efficiency profile with respect to that associated with the \BpmtoDKpm and \BpmtoDpipm sample. They are identical to the requirements listed in Sect.~\ref{sec:selection} where possible; the requirements on variables used to train the BDT follow those described in Ref.~\cite{LHCbGGSZ1fb}.

Candidate \BztoDstmu events are selected using information from the muon detector systems.
 These events are first required to pass the hardware trigger
  which selects muons with a transverse momentum $\pt>1.48\gevc$. Approximately $95\%$ of the final \BztoDstmu sample is collected with this algorithm, and the remainder pass a hardware trigger which selects \Dz candidates that leave a high transverse energy deposit in the hadronic calorimeter. 
In
  the software trigger, at least
  one of the final-state particles is required to have both
  $\pt>0.8\gevc$ and impact parameter greater than $100\mum$ with respect to all
  of the PVs in the
  event. Finally, the tracks of two or more of the final-state
  particles are required to form a vertex that is significantly
  displaced from the PVs.

To reduce combinatorial background,
all charged decay products are required to be inconsistent with originating from the PV, and 
the momentum vectors of the \KS, \Dz and \B are required to be aligned with the vector between their production and decay vertices. 
The \B candidate vertex is required to be well separated from the PV in order to discriminate between $B$ decays and prompt charm decays.

The \B decay chain is refitted~\cite{Hulsbergen:2005pu} to determine the distribution of candidates across the Dalitz plot. Unlike the refit performed for \BpmtoDhpm candidates, the fit constrains only the \Dz and \KS candidates to their known masses as the \B candidate is not fully reconstructed in the semileptonic decay mode. An additional fit, in which only the \KS mass is constrained, is performed to improve the \Dz and \Dstarpm mass resolutions for use in the invariant mass fit used to determine signal yields.  

Additional requirements are included to remove $\Dz\to \pip\pim\pip\pim$ decays and charmless $B$ decays, and PID criteria are placed on the kaons in $\Dz \to \KsKK$. The requirements are the same as those applied to the \BpmtoDKpm and \BpmtoDpipm candidates described in Sect.~\ref{sec:selection}. The \KS candidate mass is required to be within $20\mevcc$ of the known value~\cite{PDG2012}, and the invariant mass sum of the \Dstarpm and muon, determined using the refit containing the \Dz and \KS mass constraints, is required to be less than $5000\mevcc$.

\begin{figure}[htbp]
\begin{center}
\includegraphics[width=0.49\textwidth]{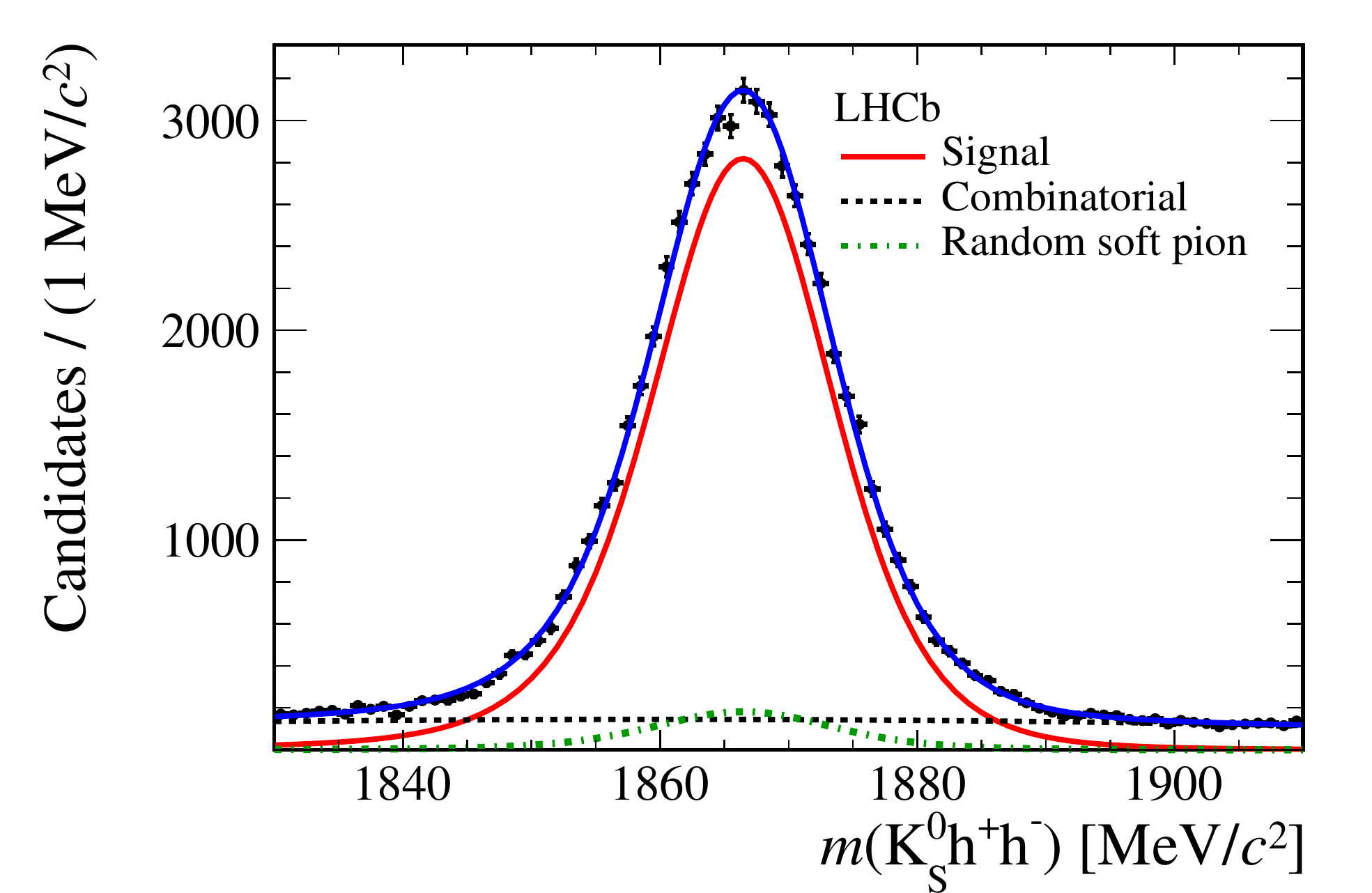} 
\includegraphics[width=0.49\textwidth]{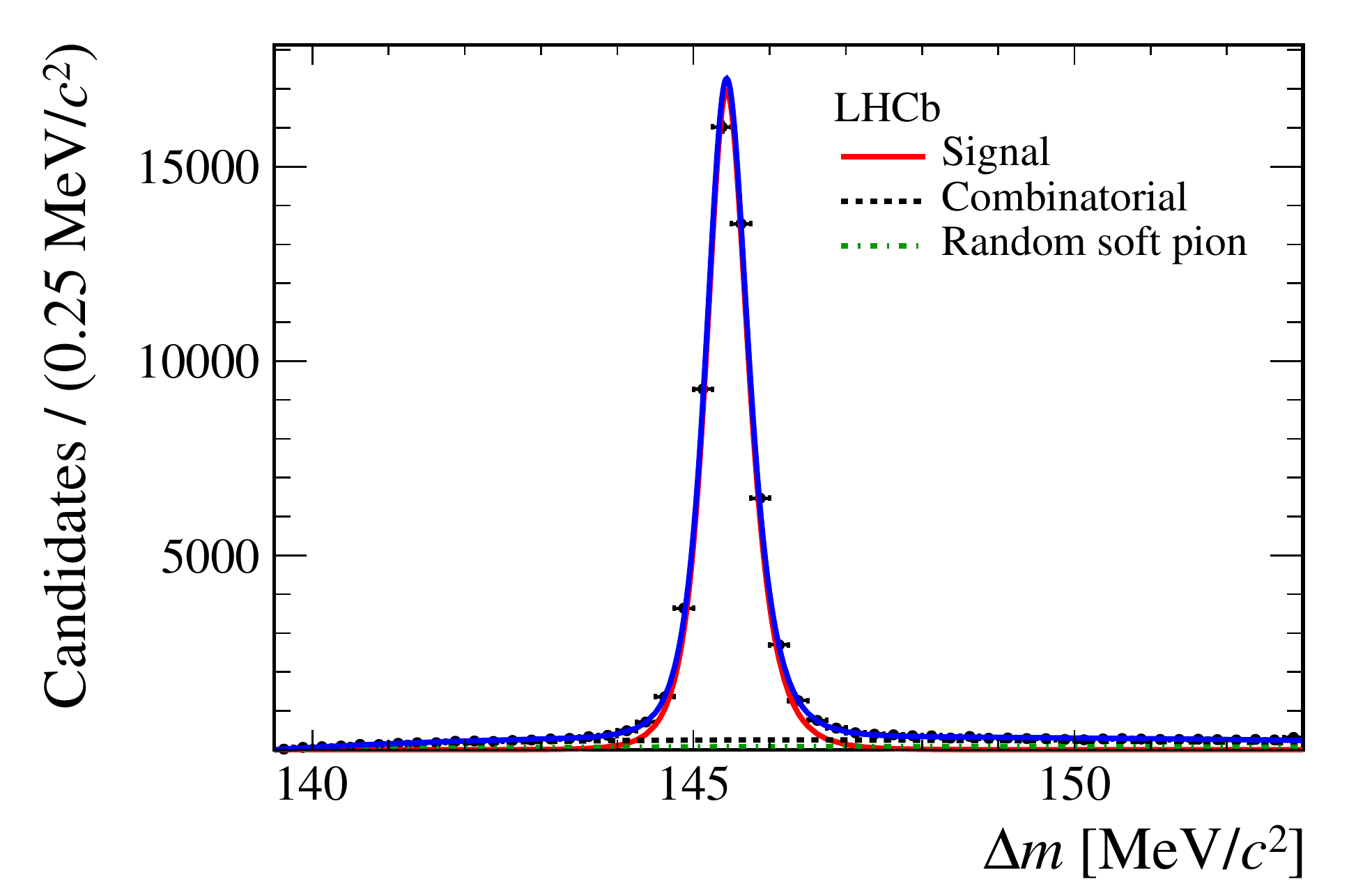} 
\caption{Result of the simultaneous fit to \BztoDstmu, $\Dstar\pm\to\DorDbar^0(\to\KsPiPi)\pipm$ decays with downstream \KS candidates, in 2012 data. A two-dimensional fit is performed in (left) $m(\Kshh)$ and (right) $\Delta m$. The (blue) total fit PDF is constructed from (solid red) signal, (dashed black) combinatorial background and (dotted green) random soft pion background.}
\label{fig:slfit2012pipiDD}
\end{center}
\end{figure}

The candidate \Dz invariant mass, $m(\Kshh)$, and the invariant mass difference $\Delta m \equiv m(\Kshh\pipm)-m(\Kshh)$ are fitted simultaneously to determine the signal yields. No significant correlation between these two variables is observed. This two-dimensional parameterisation allows the yield of selected candidates to be measured in three categories: true $\Dstarpm$ candidates (signal), candidates containing a true \Dz but random soft pion (RSP) and candidates formed from random track combinations that fall within the fit range (combinatorial background). An example projection is shown in Fig.~\ref{fig:slfit2012pipiDD}. The result of a two-dimensional extended, unbinned, maximum likelihood fit is superimposed. The fit is performed simultaneously for the two \Dz final states and the two \KS categories with some parameters allowed to be independent between categories. Candidates selected from data recorded in 2011 and 2012 are fitted separately, due to their slightly different Dalitz plot efficiency profiles. The fit range is $1830 < m(\Kshh) < 1910\mevcc$ and $139.5 < \Delta m < 153.0\mevcc$. The $m(\Kshh)$ range is chosen to be within a region where the $\Delta m$ resolution does not vary significantly.

The signal is parameterised in $\Delta m$ with a sum of two modified Gaussian PDFs, each given by
\begin{equation}
f(\Delta m ; \mu_m, \sigma, \beta, \delta) = \frac{\delta \exp \left(-0.5 \left[\beta + \delta\log \left( \frac{\Delta m - \mu_m}{\sigma}  + \sqrt{1 + \left ( \frac{\Delta m - \mu_m}{\sigma} \right )^2} \right) \right]^2 \right)}{ \sigma \sqrt{2\pi [1 + \left( \frac{\Delta m - \mu_m}{\sigma} \right)^2]}},
\end{equation}
where $\mu_m$, $\sigma$, \Pbeta and \Pdelta are floating parameters in the fit. The parameter $\mu_m$ is shared in all data categories and the remaining parameters are fitted separately for long and downstream \KS candidates. The combinatorial and RSP backgrounds are both parameterised with an empirical model given by
\begin{equation}
f(\Delta m ; \Delta m_0, x, p_1, p_2) =
\left [1 - \exp\left (-\frac{\Delta m - \Delta m_0}{x}\right )\right ] \left({\frac{\Delta m}{\Delta m_0}}\right )^{p_1} + p_2\left (\frac{\Delta m}{\Delta m_0} -1\right)
\end{equation}
for $\Delta m - \Delta m_0 > 0$ and $f(\Delta m)=0$ otherwise,
where $\Delta m_0$, $x$, $p_1$ and $p_2$ are floating parameters. The parameter $\Delta m_0$, which describes the kinematic threshold for a $\Dstarpm \to \DorDbar{}^0 \pi^\pm$ decay, is shared in all data categories, and for both the combinatorial and RSP shapes. The remaining parameters are determined separately for \KsPiPi and \KsKK candidates. 

The signal and RSP PDFs in $m(\Kshh)$ are described by Eq.~\ref{eq:cruijff}, where $m_0$, $\sigma$, $\alpha_L$, and $\alpha_R$ are all free parameters. All of the parameters in the signal and RSP PDFs are constrained to be the same since both describe a true \Dz candidate, but the parameters are fitted separately for \KsPiPi and \KsKK, due to the different phase space available in the \Dz decay. The combinatorial background is parameterised by a second-order polynomial. 

In total a sample with a signal yield of $123\,600$ candidates is selected. 
The size of the sample is approximately 40 times larger than the \BtoDK yield. The signal mass range is defined as 1840--1890 (1850--1880)\mevcc in $m(\KsPiPi)$ ($m(\KsKK)$) and 143.9--146.9\mevcc in $\Delta m$. Within this range the background components account for 3--6\% of the yield depending on the category.

\section{Determining the \texorpdfstring{$F_i$}{Fi} fractions from the semileptonic control channel}
\label{sec:corr}
\begin{figure}[tb]
\centering
\includegraphics[width=0.47\textwidth, trim=0 0 0 2cm, clip=true]{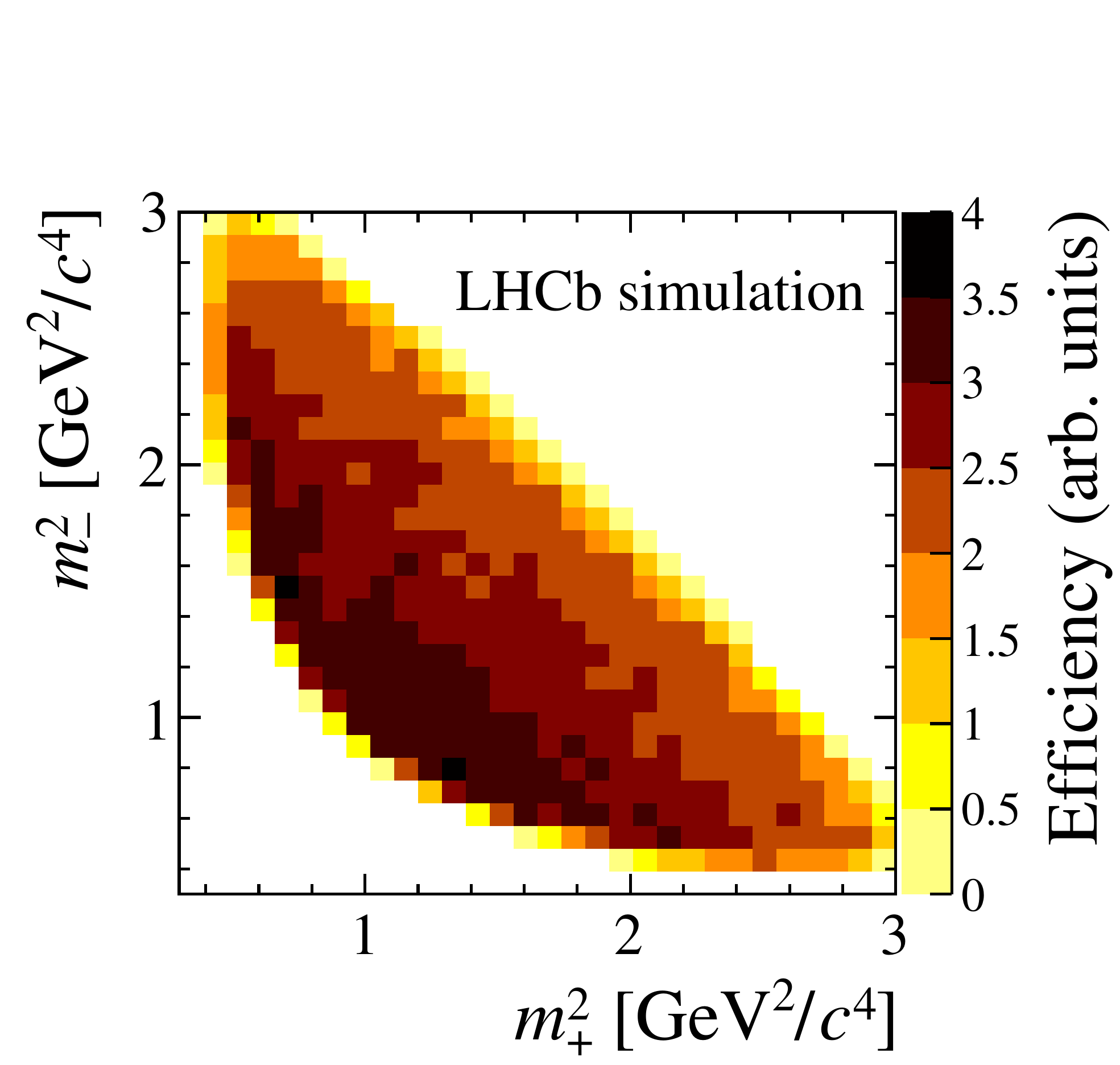}\hfill
\includegraphics[width=0.47\textwidth, trim=0 0 0 2cm, clip=true]{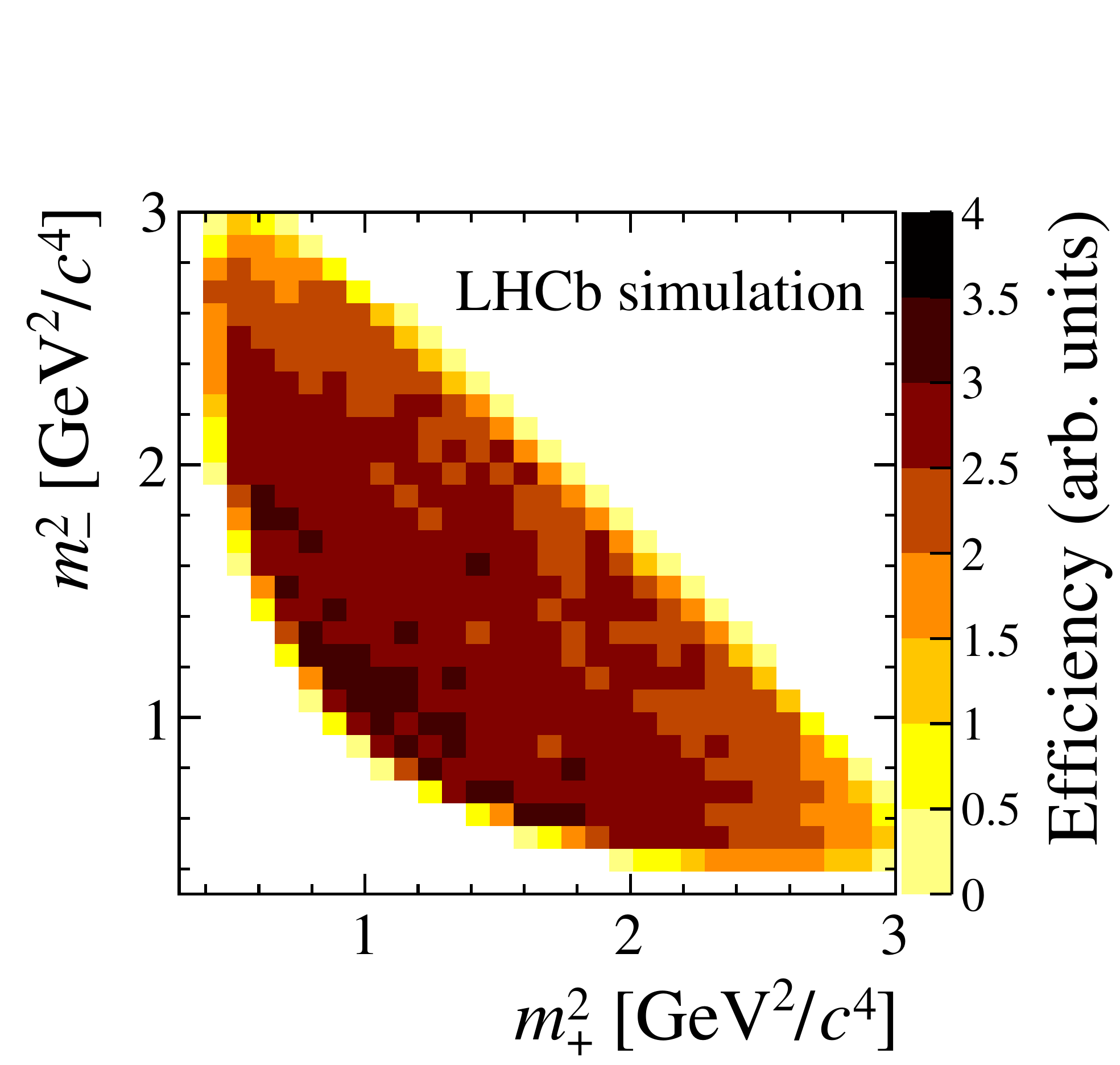}
\caption{Example efficiency profiles of (left) \BtoDpi and (right) \BztoDstmu decays in simulation. These plots refer to downstream \KS candidates under 2012 data taking conditions.}
\label{fig:mceff}
\end{figure}

The two-dimensional fit in $m(\Kshh)$ and $\Delta m$ of the \BztoDstmu decay is repeated in each Dalitz plot bin, resulting in a raw control decay yield, $R_i$, for each bin $i$. Due to the differences in the efficiency profile over the Dalitz plot between $D$ mesons originating from the control decay \BztoDstmu and those originating from the signal decay \BtoDK, the measured relative proportions of the $R_i$ values are not equivalent to the $F_i$ fractions required to determine the \CP parameters. The differences in the efficiency profiles, which originate from the different selections of the candidates from the signal and control decay modes, must be corrected for. The efficiency profiles from simulation of \DtoKsPiPi decays are shown in Fig.~\ref{fig:mceff}. They show a variation of approximately 50$\%$ between the highest and lowest efficiency regions. The variation over the \DtoKsKK Dalitz plot is 35$\%$. As the individual Dalitz plot bins cover regions of different efficiency the variation from the Dalitz plot bin with the highest efficiency and that with the lowest is approximately 30$\%$ (15$\%$) for \DtoKsPiPi (\DtoKsKK) decays.

To understand the differences between the efficiency profiles of \BztoDstmu and \BtoDK decays, 
we compare the distributions of \BztoDstmu and \BtoDpi observed in data and simulation. 
The reason for choosing \BtoDpi is that the efficiency profile is the same as for \BtoDK (as verified in simulation), but the channel has higher yields than \BtoDK. Moreover the \BtoDpi has a level of interference, and hence \CP violation, that is expected to be an order of magnitude smaller than in \BtoDK, allowing the differences in efficiency profiles to be separated from differences arising from interference effects.  The yield of \BtoDpi candidates in each bin is determined by fitting the invariant mass spectrum of candidates in each bin using the parameterisation determined in Sect.~\ref{sec:selection}.

\begin{figure}[tb]
\centering
\includegraphics[width=0.49\textwidth]{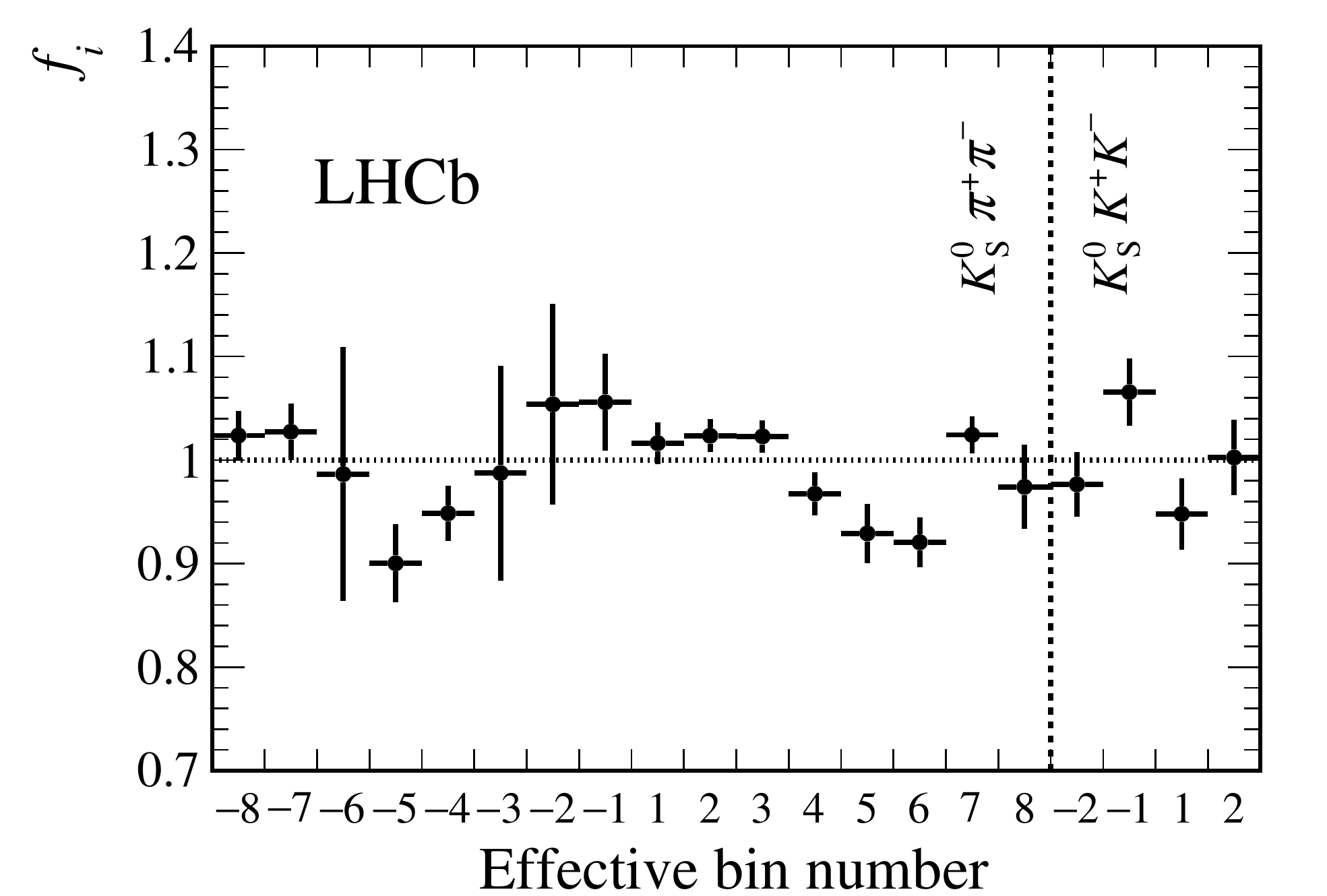}
\includegraphics[width=0.49\textwidth]{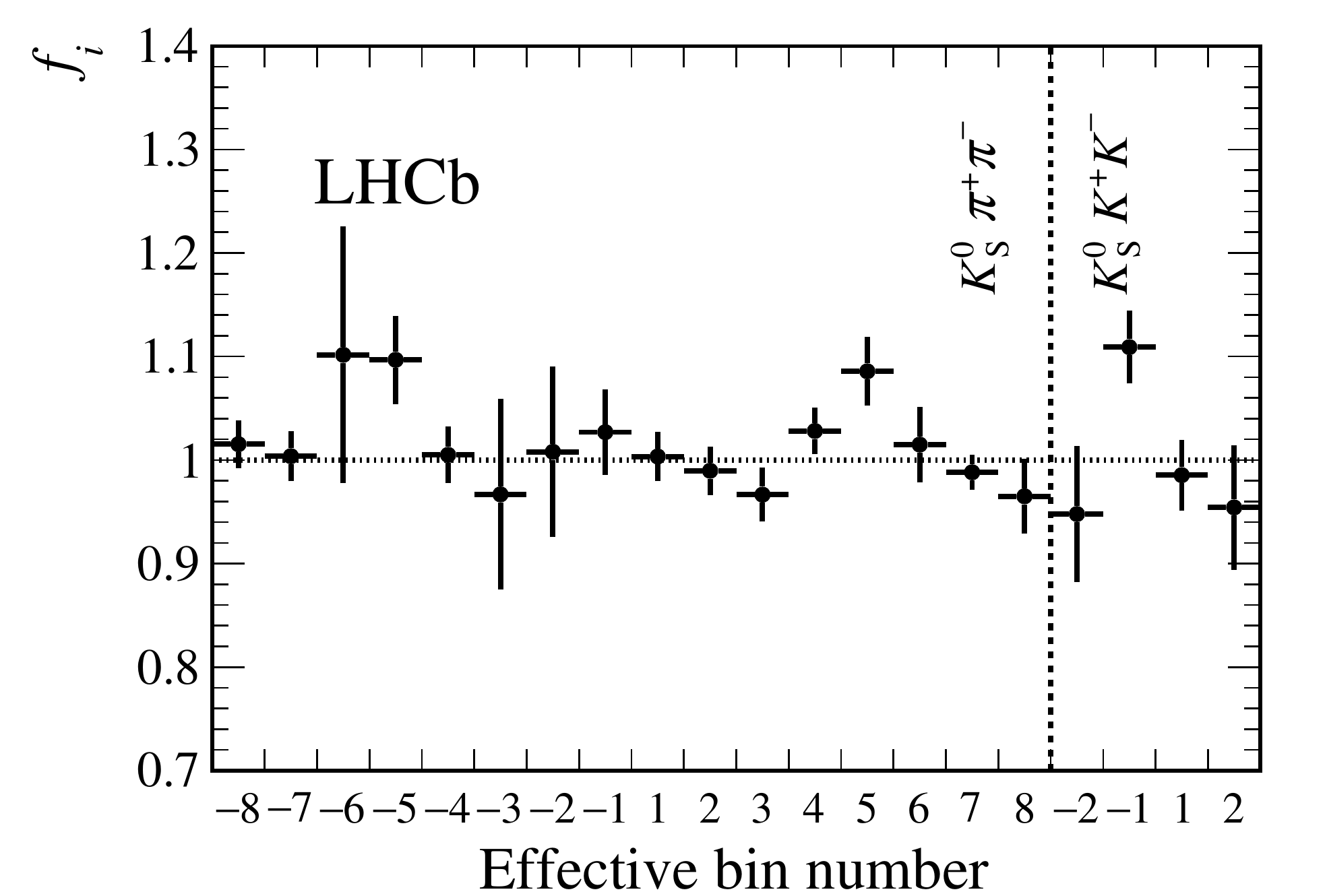}
\caption{Fractional yield ratios of \BtoDpi to \BztoDstmu decays, $f_i$, for each effective Dalitz plot bin. The vertical dashed line separates the ratios for \KsPiPi (bins $-8$ to 8) and \KsKK (bins $-2$ to 2). The left (right) plot shows the values of $f_i$ before (after) correcting for the efficiency differences.
\label{fig:before}}
\end{figure}

Figure~\ref{fig:before} shows the ratio of fractional signal yields, $f_i$, between \BtoDpi and \BztoDstmu in each Dalitz plot bin. The ratios are averaged over the two \KS samples and two periods of data taking in different experimental conditions. To increase the sample size in each bin, the yield in bin $i$ for \Dzb events is combined with the yield in bin $-i$ for \Dz events. Where the combination of yields is taken in this manner, exploiting the symmetry of the Dalitz plot, the bin number is referred to as the \emph{effective} bin. Differences of up to 10$\%$ from unity are observed in the values of $f_i$. These cannot be explained by the small amount of \CP violation in \BtoDpi decays which is expected to vary the fractional yields by 3\% or less, on the assumption that the magnitude of the interference in \BtoDpi decays is $r_B^{\pi} = 0.01$.

The raw yields of the control decay must therefore be corrected to take into account the differences in efficiency profiles. For each Dalitz plot bin a correction factor is determined,
\begin{equation}
\xi_{i} \equiv \frac{ \int_{\mathcal{D}_i} \etaDK\, |A|^2\, d\mathcal{D}}{ \int_{\mathcal{D}_i} \etaDst\, |A|^2\,d\mathcal{D}},
\end{equation}
where \etaDK and \etaDst are the efficiency profiles of the \BtoDK and \BztoDstmu decays, respectively, and $|A|^2$ is the Dalitz plot intensity for the \Dz decay. The amplitude models used to determine the Dalitz plot intensity for the correction factor are those from Ref.~\cite{BABAR2008} and Ref.~\cite{BABAR2010} for the \KsPiPi and \KsKK decays, respectively. The amplitude models used here only provide a description of the intensity distribution over the Dalitz plot and introduce no significant model dependence into the analysis. The simulation is used to determine the efficiency profiles \etaDK and \etaDst. The simulations are generated assuming a flat distribution across the \Kshh phase space; hence the distribution of simulated events after triggering, reconstruction and selection is directly proportional to the efficiency profile. The correction factors are determined separately for data reconstructed with each \KS type as the efficiency profile is different between the two \KS categories. 

The $F_i$ values can be determined via the relation $F_i = h'\xi_iR_i$, where $h'$ is a normalisation factor such that the sum of all $F_i$ is unity. The total uncertainty on $F_i$ is a combination of the uncertainty on $R_i$ due to the size of the control channel, and the uncertainty on $\xi_i$ due to the limited size of the simulated samples. The two contributions are similar in size.

To check the effect of the correction, the resulting $F_i$ values are compared to the observed population as a function of the Dalitz plot in \BtoDpi data.  Figure~\ref{fig:before}, showing the ratio of \BtoDpi fractional yields to raw \BztoDstmu fractional yields, gives a \chisq per degree of freedom ($\chisqndf$) of $48.1/20$ when considering the deviation from unity. When the corrected \BztoDstmu yields are used the fit quality improves to $\chisqndf = 29.5/20$ as seen in Fig.~\ref{fig:before}. Although the \chisq is calculated with respect to unity, the true value of $F_i$ in each bin has a variation of order $3\%$ due to \CP violation in the \BtoDpi decay.

\section{Dalitz plot fit}
\label{sec:analysis}

The Dalitz plot fit is used to measure the \CP-violating parameters $x_{\pm}$ and $y_{\pm}$, as introduced in Sect.~\ref{sec:formalism}.  Following Eq.~\ref{eq:populations2}, these parameters can be determined from the populations of each 
$B^\pm \to D K^\pm$ Dalitz plot bin, given the external information from the $c_i$, $s_i$ parameters from CLEO-c and the values of $F_i$ from the semileptonic control decay modes. 
 
Although  the absolute numbers of $B^+$ and $B^-$  decays integrated over the Dalitz plot have some dependence on $x_{\pm}$ and $y_{\pm}$, the sensitivity gained compared to using just the relative bin-to-bin yields is negligible. Consequently the integrated yields are not used and the analysis is insensitive to charged $B$ meson production and detection asymmetries. The observed size of the asymmetry of the integrated yields is consistent with that expected from the production and detection asymmetries, and the dependence on $x_{\pm}$ and $y_{\pm}$.  

A simultaneous fit is performed on the \BpmtoDhpm data, which are further split into the two $B$ charges, the two \KS categories, the $\Bpm \to D \Kpm$ and $\Bpm \to D \pipm$ candidates, and the two \DtoKshh  final states. The \BtoDK and \BtoDpi samples are fitted simultaneously because the yield of \BtoDpi signal in each Dalitz plot bin is used to determine the yield of misidentified events in the corresponding \BtoDK Dalitz plot bin.   
The PDF parameters for both the signal and background invariant mass distributions are fixed to the values determined in the invariant mass fit described in Sect.~\ref{sec:selection}. The $B$ mass range is reduced to 5150--5800\gevcc to reduce systematic uncertainties from the partially reconstructed background.  The yields of all background contributions in each bin are free parameters, apart from the yields in bins in which an auxiliary fit determines the yield to be negligible. These are set to zero to facilitate the calculation of the uncertainty matrix.  The yields of signal candidates for each bin in the $\Bpm\to D \pipm$ sample are also free parameters.   The amount of signal in each bin for the $\Bpm\to D \Kpm$ sample is determined by varying the integrated yield over all Dalitz plot bins and the  $x_{\pm}$ and $y_{\pm}$ parameters. In the fit the values of $F_i$ are Gaussian-constrained within their uncertainties. The values of $c_i$ and $s_i$ are fixed to their central values. In order to assess the impact of the  \DtoKsKK data, the fit is then repeated including only the \DtoKsPiPi sample.

A large ensemble of pseudo-experiments is performed to validate the fit procedure.  In each pseudo-experiment the numbers and distribution of signal and background candidates are generated according to the expected distribution  in data, and the full fit procedure is then executed.  The input values for $x_{\pm}$ and $y_{\pm}$ are set close to those determined by previous measurements~\cite{HFAG}.   The uncertainties estimated by the fit are consistent with the size of the uncertainties estimated by the pseudo-experiments.   However, small biases, with sizes around $10\%$ of the statistical uncertainty, are observed in the central values. These biases  
are due to the low event yields in some of the bins and are observed to reduce in simulated experiments of larger size. The central values are corrected for the biases.

The results of the fits are presented in Table~\ref{tab:theresults}.    The statistical uncertainties are compatible with those predicted by the simulated pseudo-experiments.   The systematic uncertainties are discussed in Sect.~\ref{sec:syst}. The inclusion of \DtoKsKK data improves the precision on $x_{\pm}$ by around 10\% and by a smaller amount for $y_{\pm}$.
This is expected, as the measured values of $c_i$ in this decay, which multiply $x_\pm$ in Eq.~\ref{eq:xydefinitions}, are significantly larger than those of $s_i$, which multiply $y_\pm$~\cite{CLEOCISI}.

\begin{table}[tb]
\centering
\caption{Results for $x_{\pm}$ and $y_{\pm}$ from fits of
both  the $D \to \KS \pi^+\pi^-$ and $D \to \KS K^+K^-$ samples, and from fits of 
the $D \to \KS \pi^+\pi^-$ sample only.  The first, second, and third uncertainties are the statistical, the experimental systematic, and the error associated with the precision of the strong-phase parameters, respectively.
}
\label{tab:theresults} \vspace*{0.1cm}
\begin{tabular}{r|rr} 
Parameter &  \multicolumn{1}{c}{All data} &  \multicolumn{1}{c}{$D \to \KS  \pi^+\pi^- $ only} \\ \hline
$x_+$ [$\times 10^{-2}$] &   $-7.7 \pm 2.4 \pm 1.0 \pm 0.4$  &  $-7.5 \pm 2.7 \pm 1.1 \pm 0.5$ \\
$x_-$ [$\times 10^{-2}$] &   $2.5 \pm 2.5 \pm 1.0 \pm 0.5$   &  $2.6 \pm 2.8 \pm 1.1 \pm 0.7$ \\
$y_+$ [$\times 10^{-2}$] &   $-2.2 \pm 2.5 \pm 0.4 \pm 1.0$  &  $-1.4 \pm 2.6 \pm 0.6 \pm 0.9$ \\
$y_-$ [$\times 10^{-2}$] &   $7.5 \pm 2.9 \pm 0.5 \pm 1.4$   &  $7.5 \pm 3.0 \pm 0.4 \pm 1.5$ \\
\end{tabular}
\end{table}

\begin{figure}[tb]
\begin{center}
\includegraphics[width=0.48\textwidth]{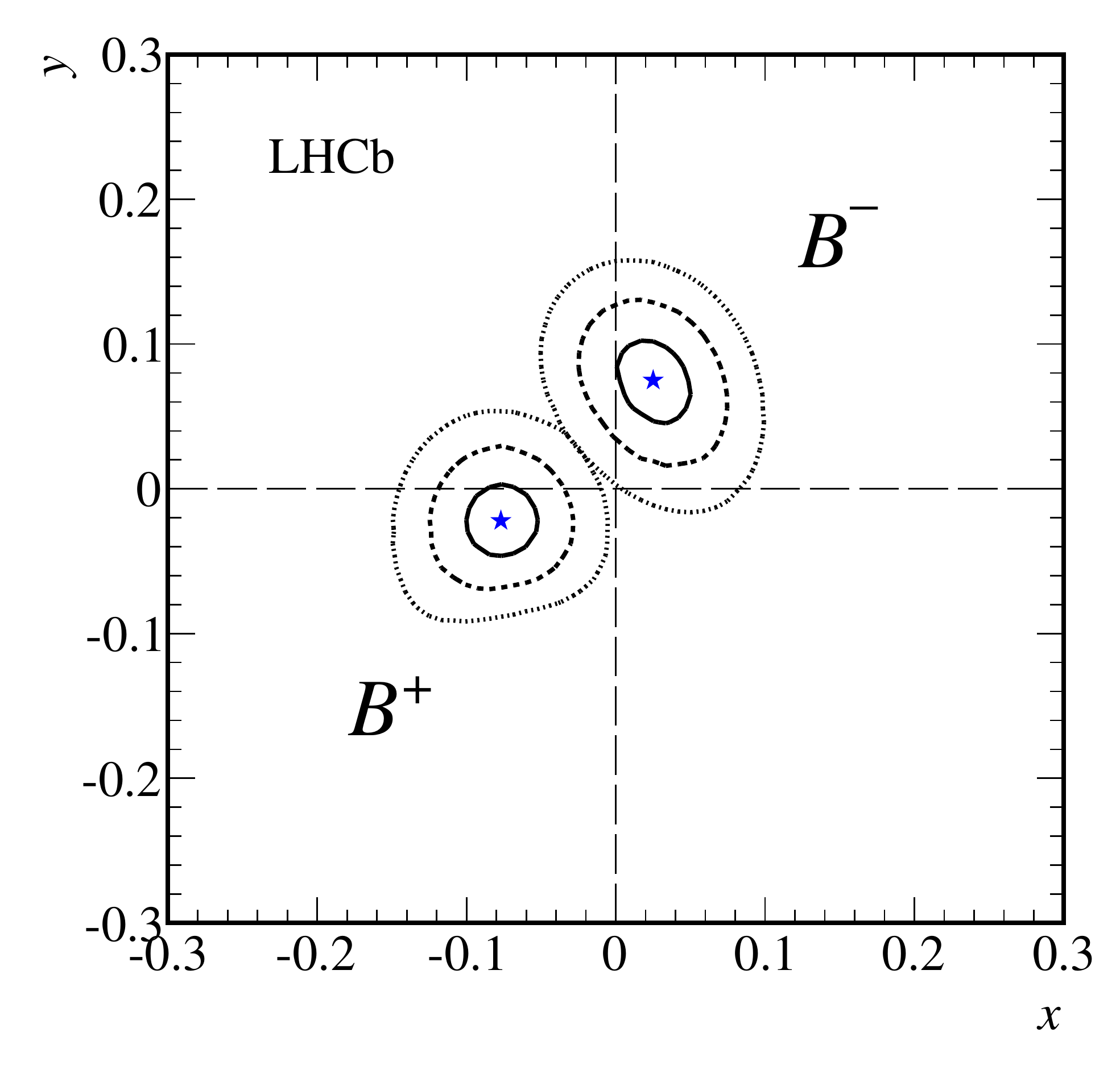}
\caption{\small Confidence levels at 39.3\%, 86.5\% and 98.9\% probability for 
$(x_+,y_+)$ and $(x_{-}, y_{-})$ as measured in $B^\pm \to D K^\pm$ decays (statistical uncertainties only). The parameters $(x_+,y_+)$ relate to \Bp decays and $(x_{-}, y_{-})$ refer to \Bm decays. 
The stars represent the best fit central values.}
\label{fig:sunnysideup}
\end{center}
\end{figure}

The measured values of $(x_{\pm}, y_{\pm})$ from the fit to all data, with their likelihood contours corresponding to statistical uncertainties only, are displayed in Fig.~\ref{fig:sunnysideup}. 
The expected signature for a sample that exhibits \CP violation is that the two vectors defined by the coordinates $(x_-,y_-)$ and $(x_+,y_+)$ should both be non-zero in magnitude and have a non-zero opening angle, which is equal to $2\gamma$. 

\begin{figure}[tb]
\centering
\includegraphics[width=0.48\textwidth]{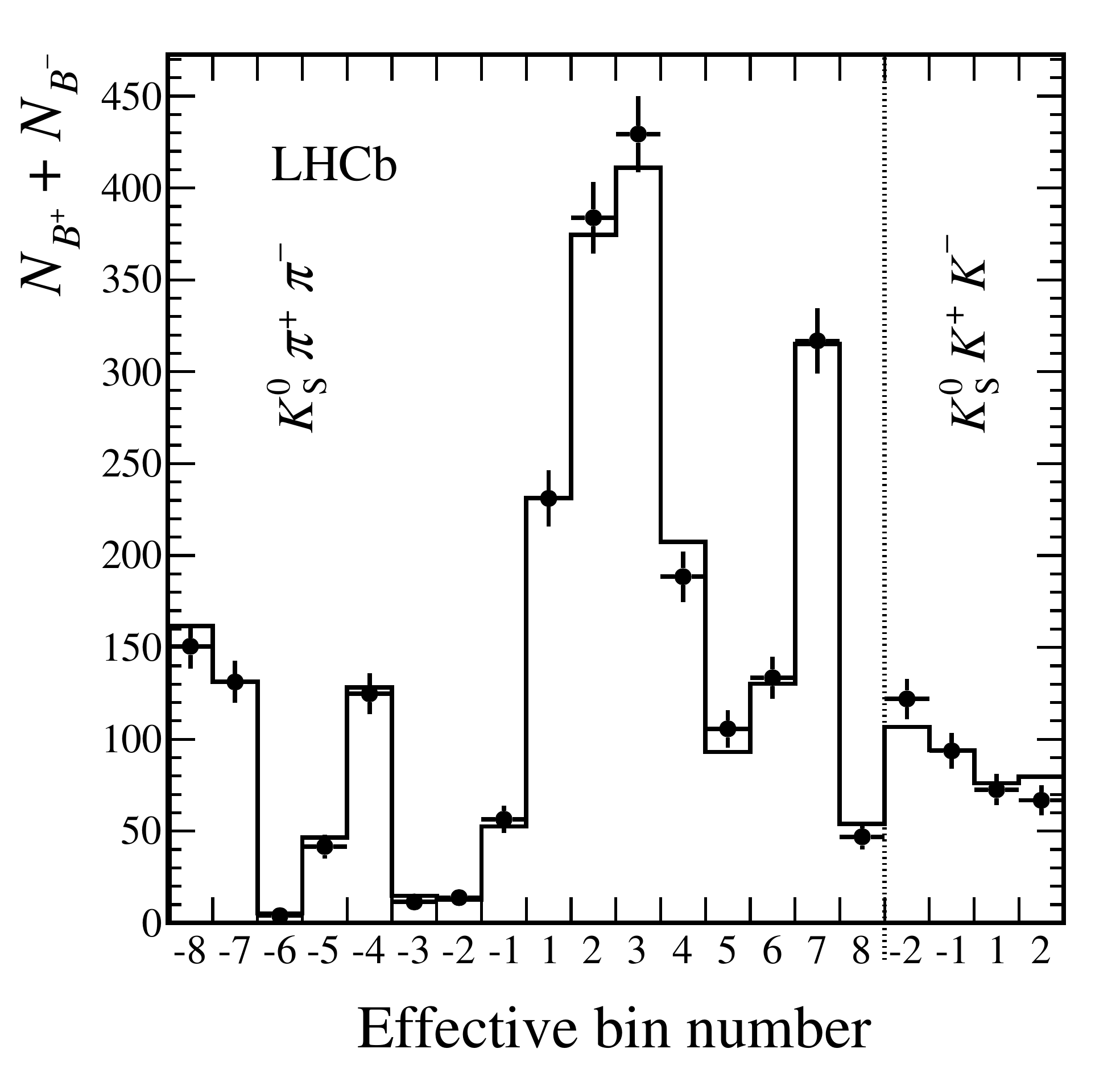}
\includegraphics[width=0.48\textwidth]{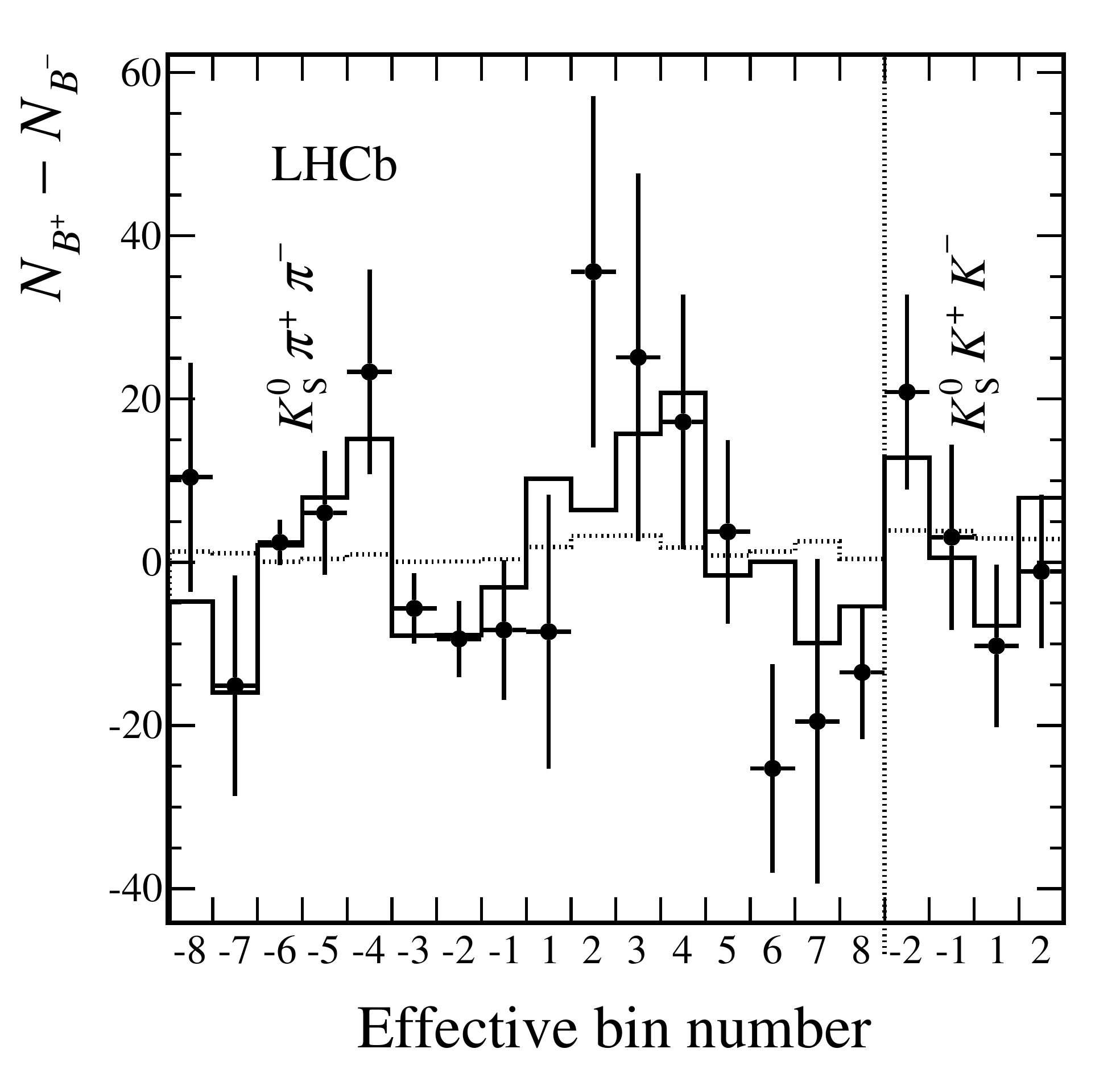}
\caption{\small 
Combinations of signal yields (data points) in effective bins compared with prediction of $(x_\pm, y_\pm)$ fit (solid histogram) for \DtoKsPiPi and \DtoKsKK decays. The dotted histograms give the prediction for $x_\pm=y_\pm=0$. The left plot shows the sum of $B^+$ and $B^-$ yields and the right plot shows the difference of $B^+$ and $B^-$ yields.}
\label{fig:bin_plots}
\end{figure}

To investigate whether the binned fit gives an adequate description of the data, a study is performed to compare the expected signal yield in each bin, given by the fitted total yield and the values of $x_\pm$ and $y_\pm$, and the observed number of signal candidates in each bin. The latter is determined by fitting directly in each bin for the $\BpmtoDKpm$ candidate yield. This study is performed using effective bin numbers and with long and downstream \KS decays combined. 
Figure~\ref{fig:bin_plots} shows the results separately for the sum of $B^+$ and $B^-$ candidates, $N_{B^+} + N_{B^-}$, and for the difference, $N_{B^+} - N_{B^-}$, which is sensitive to \CP violation.  
The expected signal yields assuming \CP symmetry ($x_\pm = y_\pm = 0$) in the $N_{\Bp}-N_\Bm$ distribution are also shown.  
These are not constant at $N_{B^+} -  N_{B^-} =0$ because they are calculated using the total $B^+$ and $B^-$ yields, which do not have identical values. 
The data and fit expectations are compatible for both distributions yielding a $\chi^2$ probability ($p$-value) of 93\% for $N_{B^+} + N_{B^-}$ and 80\% for 
$N_{B^+} - N_{B^-}$. The results for the $N_{B^+} - N_{B^-}$ distribution are less compatible with the hypothesis of \CP symmetry, which has a $p$-value of 4\%.

\section{Systematic uncertainties}
\label{sec:syst}

Systematic uncertainties are evaluated for the fits to the full data sample and are presented in Table~\ref{tab:syst_all}. 
The uncertainties arising from the CLEO-c measurements are kept separate from the other experimental uncertainties. 
\begin{table}[tb]
\centering
\caption{\small Summary of statistical, experimental, and strong-phase, uncertainties on $x_\pm$ and $y_\pm$ in the case where
both \DtoKsPiPi and \DtoKsKK decays are included in the fit. All entries are given in multiples of $10^{-2}$.}\label{tab:syst_all} 
\vspace*{0.1cm}
\begin{tabular}{l| c c c c }
\hline
Source & $\sigma(x_+)$ & $\sigma(x_-)$ & $\sigma(y_+)$ & $\sigma(y_-)$ \\
\hline 
Statistical & 2.4 & 2.5 & 2.5 & 2.9 \\
\hline
Efficiency corrections & 0.9 & 0.9 & 0.2 & 0.2 \\
Mass fit PDFs & 0.2 & 0.2 & 0.1 & 0.2 \\
Shape of \Dpi mis-identified as \DK & 0.1 & 0.1 & 0.0 & 0.1 \\
Shape of partially reconstructed backgrounds & 0.1 & 0.3 & 0.1 & 0.2\\
$c_i$, $s_i$ bias due to efficiency & 0.0 & 0.0 & 0.1 & 0.0 \\
Migration & 0.1 & 0.1 & 0.2& 0.2\\
Bias correction & 0.2 & 0.2 & 0.2 & 0.2 \\
\hline
Total experimental & 1.0 & 1.0 & 0.4 & 0.5 \\
\hline
Strong-phase-related uncertainties & 0.4 & 0.5 & 1.0 & 1.4 \\
\hline
\end{tabular}
\end{table}

\begin{figure}[tb]
\centering
\includegraphics[width=0.47\textwidth, trim=0 0 0 2cm, clip=true]{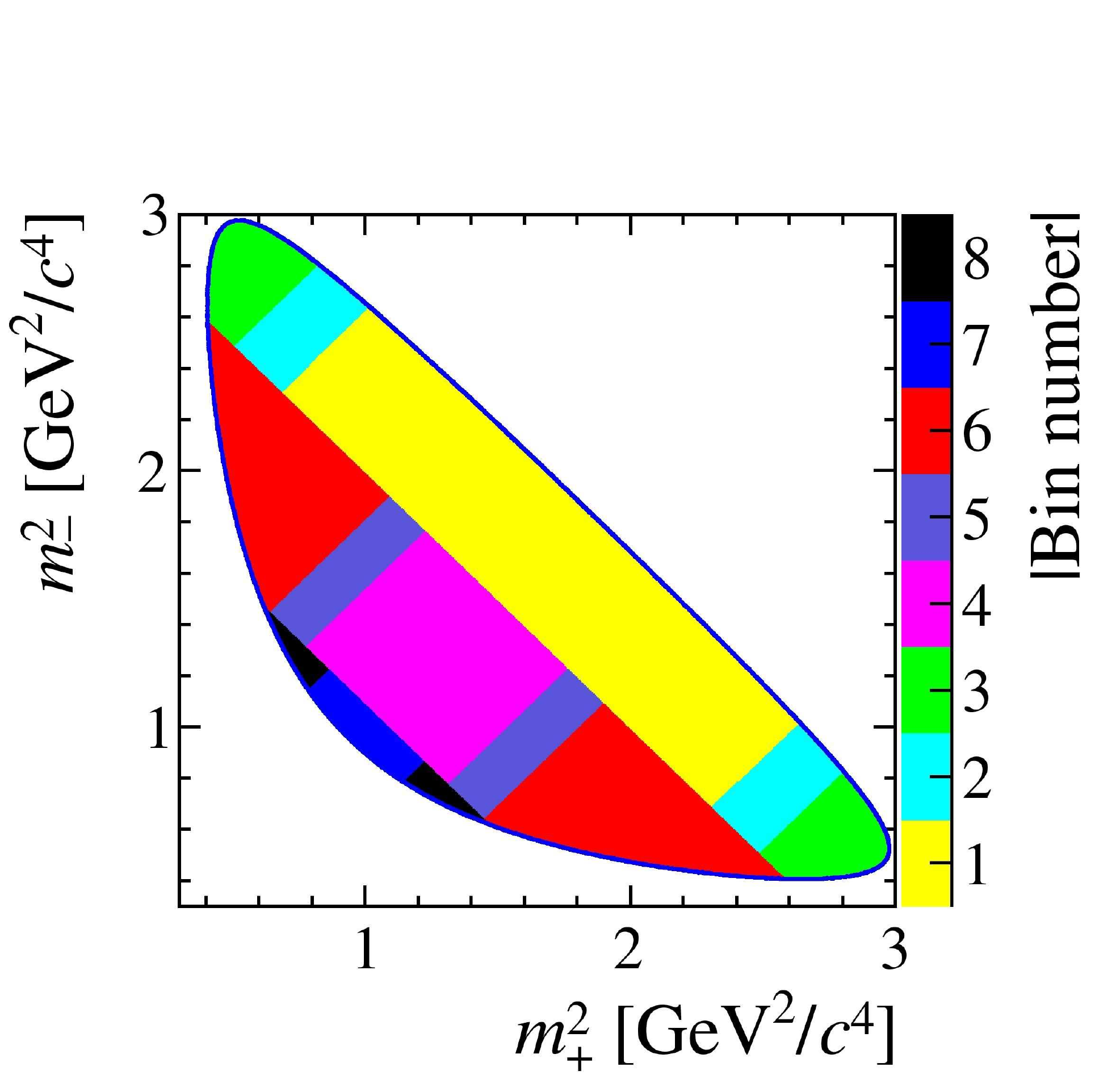}
\includegraphics[width=0.47\textwidth, trim=0 0 0 2cm, clip=true]{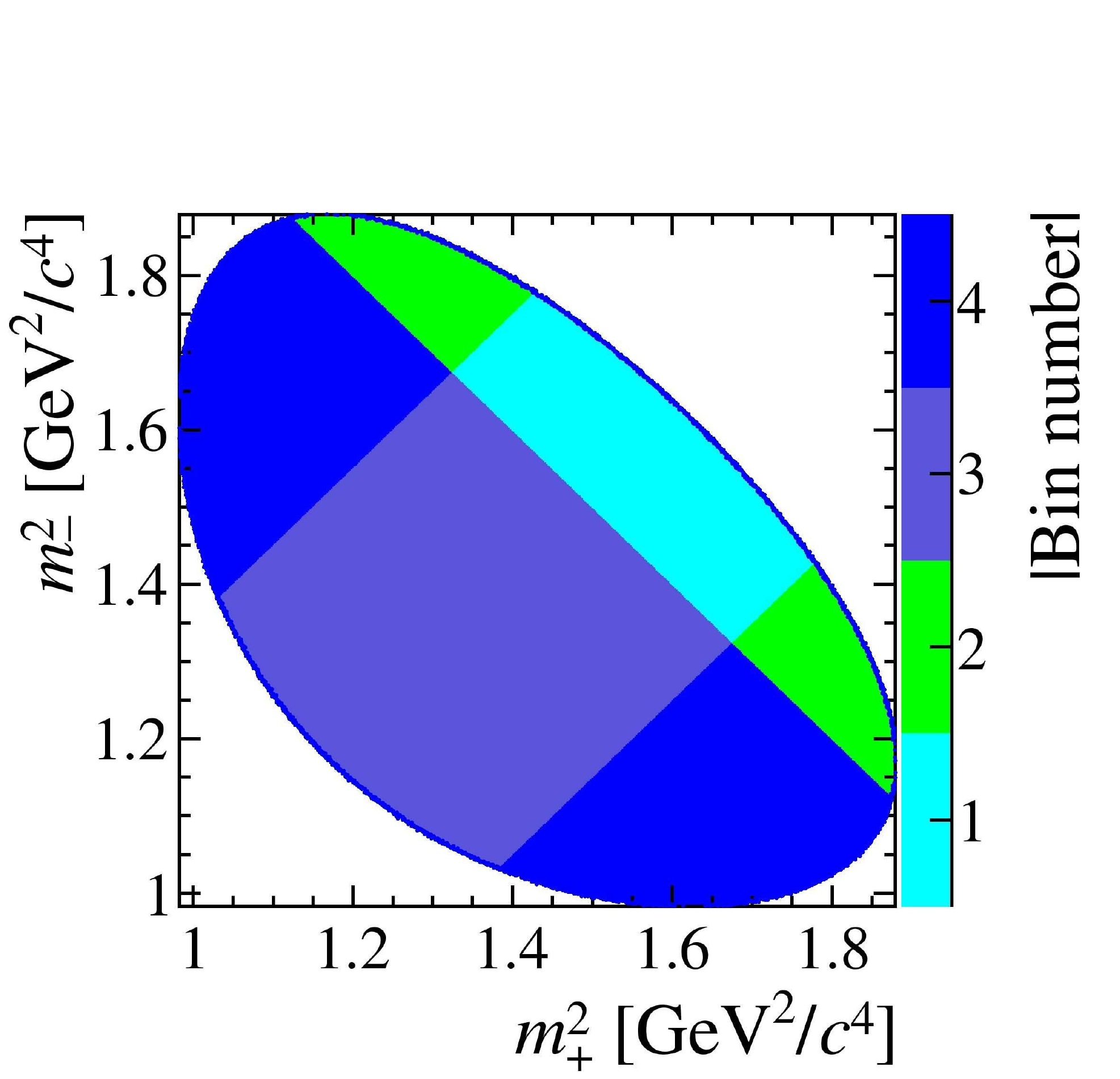}
\caption{The rectangular binning schemes for the two decays. On the left (right) is plotted the scheme for the \KsPiPi (\KsKK) decay.}
\label{fig:rect}
\end{figure}

A systematic uncertainty arises due to the mismodelling in the simulation used to derive the efficiency correction used in the determination of the $F_i$ parameters. To determine the systematic uncertainty associated with this correction, an alternative set of correction factors is calculated and used to evaluate an alternative set of $F_i$ parameters. The alternative correction factors are calculated by incorporating an extra term (Eq.~\ref{eq:altcor}) determined from a new rectangular binning scheme, as shown in Fig.~\ref{fig:rect}. The bin-to-bin efficiency variation in this rectangular scheme is significantly larger than for the default partitioning and is more sensitive to imperfections in the simulated data efficiency profile. The bin sizes are chosen to keep the expected yields in each bin as similar as possible. The yields of the \BtoDpi and \BztoDstmu decays in each bin of the rectangular scheme are compared to the predictions from the amplitude model and the simulated data efficiency profile. Differences of up to 15$\%$ are observed. These differences are consistent for the two decay modes. 
The alternative correction factors, $\xi_{i}^{ \textrm{alt}}$, are calculated using the following equation:
\begin{equation}
\label{eq:altcor}
\xi_{i}^{ \textrm{alt}} =\frac{ \int_{\mathcal{D}_i} \etaDK\, |A|^2 \, C_{D\pi}\, d\mathcal{D}}{ \int_{\mathcal{D}_i} \etaDst\, |A|^2 \, C_{\Dstar\!\mu}\, d\mathcal{D}}
\end{equation}
where the $C=C(m^2_-,m^2_+)$ terms are the ratios between the predicted and observed data yields in the rectangular binning. Many pseudo-experiments are performed in which the data are generated according to the default $F_i$ but are fitted assuming that the alternative $F_i$ set are true. The overall shift in the fitted values of the \CP parameters in comparison to their input values is taken as the systematic uncertainty, yielding $0.9\times 10^{-2}$ for \xpm and $0.2\times 10^{-2}$ for \ypm.

To assign an uncertainty for the imperfections in the description of the invariant mass spectrum, three changes to the model are considered. 
Firstly, an alternate signal shape is considered that has wider resolution and longer tails. This alternate shape uses a different form of modified Gaussian and the parameters are derived from a fit to data. Secondly, the description of the partially reconstructed background is changed to a shape obtained from a fit of the PDF to simulated \DtoKshh decays. 
Finally, the parameters of the misidentified background PDF are changed to vary the tail under the signal peak as this is the part of the PDF that is least well determined. For each change, the effects on the \CP parameters are determined using many pseudo-experiments where the data are generated with the default PDFs and fitted with the alternate models. The contributions from each change are summed in quadrature and are $(0.1\text{--}0.2) \times 10^{-2}$.

Two systematic uncertainties are evaluated that are associated with the misidentified \BpmtoDpipm background in the \BpmtoDKpm sample. 
The uncertainties on the particle misidentification efficiencies are found to have a negligible effect on the measured values of \xpm and \ypm. 
It is possible that the invariant mass distribution of the misidentified background  is not uniform over the Dalitz plot, as is assumed in the fit. This can occur through kinematic correlations between  the reconstruction efficiency on the Dalitz plot of the $D$ decay and the momentum of the bachelor pion from the \Bpm decay.  Pseudo-experiments are performed with different mass shapes input according to the Dalitz plot bin and the results of simulation studies. These experiments are then fitted assuming a uniform shape, as in data. The resulting uncertainty is up to $0.1 \times 10^{-2}$ for all \CP parameters. 

The distribution of the partially reconstructed background is varied over the Dalitz plot according to the uncertainty in the composition of this background component. This results in a different invariant mass distribution in each Dalitz plot bin. An uncertainty of $(0.1\text{--}0.3) \times 10^{-2}$ is assigned to the fitted parameters in the full data fit.

The non-uniform efficiency profile over the Dalitz plot means that the values of $(c_i, s_i)$ appropriate for this analysis can differ from those measured at CLEO-c, which correspond to the constant efficiency case. This leads to a potential bias in the determination of \xpm and \ypm. The possible size of this effect is evaluated using the LHCb simulation. The Dalitz plot bins are divided into smaller cells, and the \babar amplitude model~\cite{BABAR2008,BABAR2010} is used to calculate the values of $c_i$ and $s_i$ within each cell.   These values are then averaged together and weighted by the population of each cell after efficiency losses to obtain an effective $(c_i, s_i)$ for the bin as a whole. The results are compared with those determined assuming a constant efficiency; the differences between the two sets of results are found to be small compared with the CLEO-c measurement uncertainties.  
The data are fitted multiple times, each with different $(c_i, s_i)$ values sampled according to the size of these differences, 
and the mean shifts are assigned as a systematic uncertainty.  These shifts are less than $0.1 \times 10^{-2}$ for all \CP parameters.

For both \BtoDK and \BztoDstmu decays the resolution in $m^2_+$ and $m^2_-$ of each decay is approximately 0.005\gevgevcccc for candidates with long \KS decays and 0.006\gevgevcccc for candidates with downstream \KS.  This is small compared to the typical width of a bin but net migration away from the more densely populated bins is possible.  To first order this effect is accounted for by use of the control channel, but residual effects enter due to the different distribution in the Dalitz plot of the signal events. The uncertainty due to these residual effects is determined via pseudo-experiments, in which different input $F_i$ values are used to reflect the residual migration. The size of any possible bias is found to vary between $0.1 \times 10^{-2}$ and $0.2 \times 10^{-2}$.

An uncertainty is assigned to each \CP parameter to accompany the correction that is applied  for the small  bias observed in the fit procedure.  These uncertainties are determined by performing sets of pseudo-experiments, each generated with different values of \xpm and \ypm according to the range allowed by current experimental knowledge. 
The spread in observed bias is combined in quadrature with half the correction and the uncertainty in the precision of the pseudo-experiments. This is taken as the systematic uncertainty, and is $0.2 \times 10^{-2}$ for all \CP parameters.
The effect that a detection asymmetry between hadrons of opposite charge can have on the symmetry of the efficiency of the Dalitz plot is found to be negligible. Changes in the mass model used to describe the semileptonic control sample are found to have a negligible effect on the $F_i$ values.
 
The limited precision on $(c_i, s_i)$ coming from the CLEO-c measurement induces uncertainties on $x_\pm$ and $y_\pm$~\cite{CLEOCISI}.
These uncertainties are evaluated by 
fitting the data multiple times, each with different $(c_i, s_i)$ values sampled according to their experimental uncertainties and correlations. 
The resulting width in the distribution of $x_\pm, y_\pm$ values is assigned as the systematic uncertainty. Values of $(0.4\text{--}1.4) \times 10^{-2}$ are found for the fit to the full sample. The uncertainties are smaller than those reported in Ref.~\cite{LHCbGGSZ1fb}. This is as expected since it is found from simulation studies that the ($c_i$, $s_i$) uncertainty also depends on the sample size.

Finally, several checks are conducted to assess the stability of the results.   These include repeating the fits separately for both \KS categories, for the centre-of-mass energy at which the data were collected, and for candidates passing different hardware trigger requirements. No anomalies are found, and no additional systematic uncertainties are assigned. 

The total experimental systematic uncertainty from LHCb-related sources is determined to be $1.0 \times 10^{-2}$ on $x_+$, $1.0 \times 10^{-2}$ on $x_-$, $0.4 \times 10^{-2}$ on $y_+$, and $0.5 \times 10^{-2}$ on $y_-$.   These are all smaller than the corresponding statistical uncertainties.  The dominant contribution arises from the efficiency correction method.

After taking account of all sources of uncertainty the correlation matrix between the measured \xpm, \ypm parameters for the full data set is shown in Table~\ref{tab:corrmatall}. Correlations from the statistical and strong-phase uncertainties are included but the experimental systematic uncertainties are treated as uncorrelated. The equivalent matrix for \DtoKsPiPi decays only is shown in Table~\ref{tab:corrmatKsPiPi}.  
\begin{table}[tb]
\caption{Correlation matrix between the \xpm, \ypm parameters for the full data set. \label{tab:corrmatall}}
\begin{center}
\begin{tabular}{c | c c c c }
      & $x_+$ & $x_-$ & $y_+$ & $y_-$ \\
\hline
$x_+$ & $\phantom{-} 1.000$ & $\phantom{-} 0.027$ & $\phantom{-} 0.003$ & $-0.007$ \\
$x_-$ &  & $\phantom{-} 1.000$ & $\phantom{-} 0.009$ & $-0.200$ \\
$y_+$ & & & $\phantom{-} 1.000$ & $-0.016$ \\
$y_-$ & & & & $\phantom{-} 1.000$ \\
\end{tabular}
\end{center}
\end{table}
\begin{table}[th]
\caption{Correlation matrix between the \xpm, \ypm parameters for \KsPiPi decays only. \label{tab:corrmatKsPiPi}}
\begin{center}
\begin{tabular}{c | c c c c }
      & $x_+$ & $x_-$ & $y_+$ & $y_-$ \\
\hline
$x_+$ & $\phantom{-} 1.000$ & $\phantom{-} 0.017$ & $-0.038$ & $-0.008$ \\
$x_-$ &   & $\phantom{-} 1.000$ & $\phantom{-} 0.009$ & $-0.220$ \\
$y_+$ &  &  & $\phantom{-} 1.000$ & $\phantom{-} 0.004$ \\
$y_-$ &  &  &  & $\phantom{-} 1.000$ \\
\end{tabular}
\end{center}
\end{table}

 The systematic uncertainties for the case  where only \DtoKsPiPi  decays are included are also given in Table~\ref{tab:theresults}. The total experimental systematic in this case is larger and this is primarily driven by a larger systematic effect due to the simulation-derived efficiency correction, for which the systematic uncertainty for the \DtoKsKK decays partially compensates. The uncertainties due to the CLEO-c strong-phase measurements are also slightly larger when considering only \DtoKsPiPi decays due to the dependence of this systematic uncertainty on the signal sample and its size.   

\section{Results and interpretation}
\label{sec:discussion}

The results for $x_\pm$ and $y_\pm$ can be interpreted in terms of the underlying physics parameters $\gamma$, $r_B$ and $\delta_B$.  This interpretation is performed using a frequentist approach with Feldman-Cousins ordering~\cite{FELDMANCOUSINS}, using the same procedure as described in Ref.~\cite{BELLEMODIND}, yielding confidence levels for the three physics parameters. 

\begin{figure}[tb]
\begin{center} 
\includegraphics[width=0.45\textwidth]{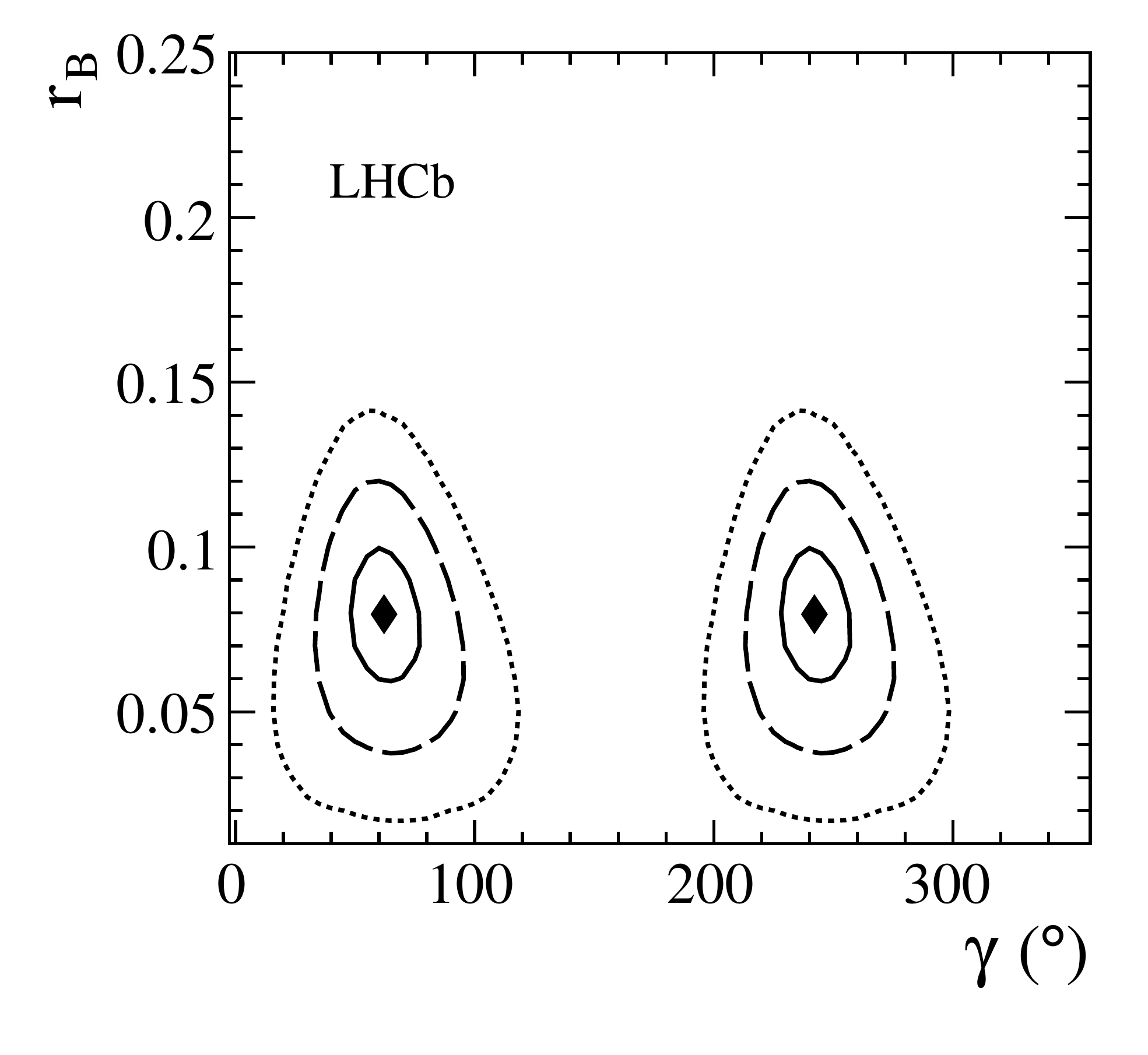}
\includegraphics[width=0.45\textwidth]{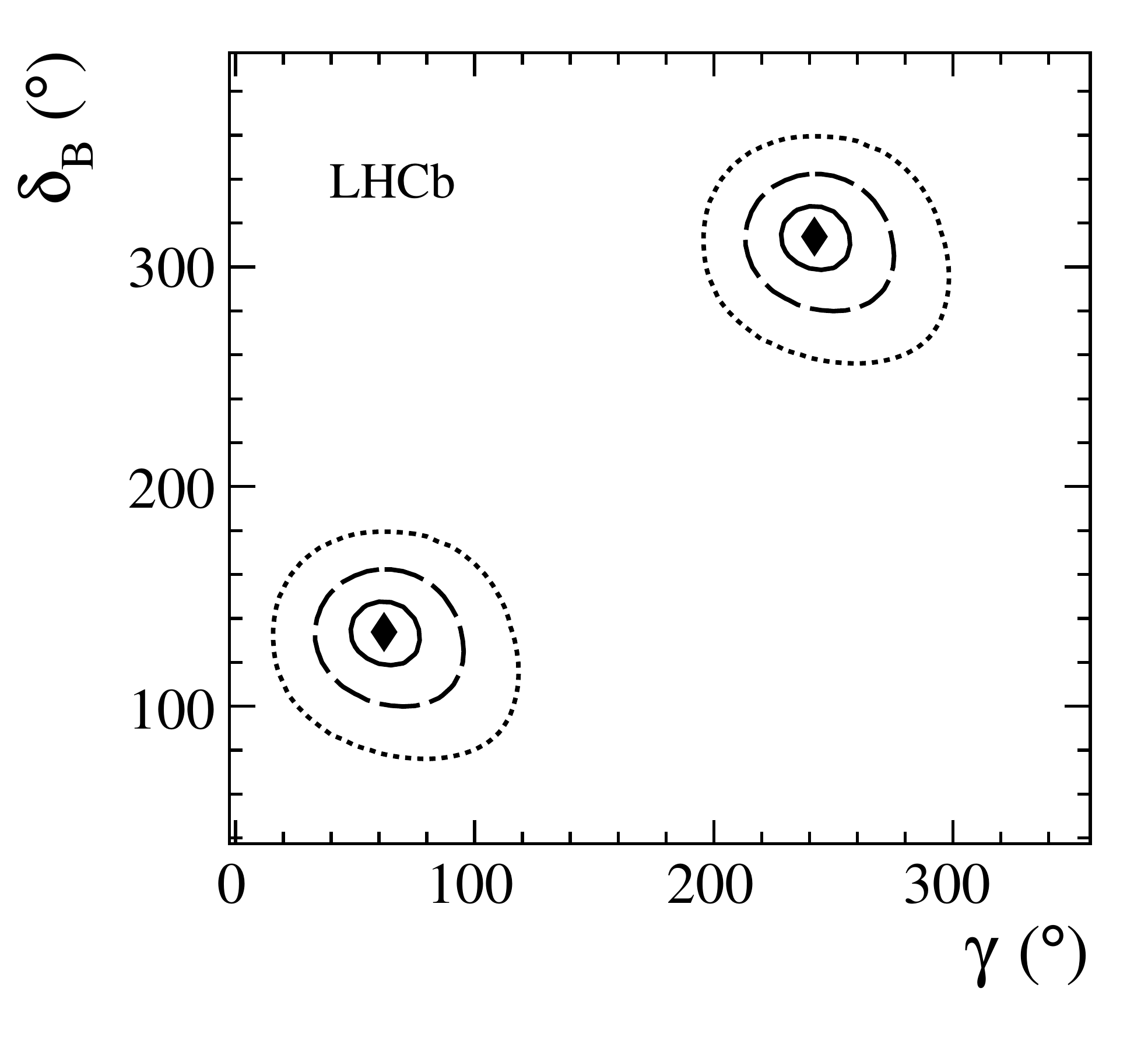}
\caption{\small The three-dimensional confidence volumes, corresponding to 19.9\%, 73.9\% and 97.1\% confidence levels, are projected onto the $(\gamma, r_B)$ and $(\gamma, \delta_B)$ planes. The confidence levels are given by solid, dashed and dotted contours. 
The diamonds mark the central values.}
\label{fig:twodscans}
\end{center}
\end{figure}

In Fig.~\ref{fig:twodscans} the projections of the three-dimensional surfaces bounding the one, two and three standard deviation volumes onto the $(\gamma, r_B)$ and $(\gamma, \delta_B)$ planes are shown. The LHCb-related systematic uncertainties are taken as uncorrelated and correlations of the CLEO-c and statistical uncertainties are taken into account. The statistical and systematic uncertainties on $x_\pm$ and $y_\pm$ are combined in quadrature.

The solution for the physics parameters has a two-fold ambiguity: $(\gamma, \delta_B) \rightarrow (\gamma + 180^\circ, \delta_B + 180^\circ)$.  Choosing the solution that satisfies $0 < \gamma < 180^\circ$  yields $r_B = 0.080^{\,+0.019}_{\,-0.021}$, $\gamma = (62^{\,+15}_{\,-14})^\circ$ and $\delta_B = (134^{\,+14}_{\,-15})^\circ$. The values for $\gamma$ and $r_B$ are consistent with the world average of results from previous experiments ~\cite{HFAG}. The significant increase in precision compared to the measurement in Ref.~\cite{LHCbGGSZ1fb} is due to a combination of increased signal yield, lower systematic uncertainties and a higher central value for $r_B$.

\section{Conclusions}
\label{sec:conclusions}

Approximately 2580 $B^\pm \to D K^\pm$ decays, with the $D$ meson decaying either to $\KS \pi^+\pi^-$ or $\KS K^+ K^-$,  are selected from data corresponding to and integrated luminosity of 3.0~${\rm fb^{-1}}$ collected by LHCb in 2011 and 2012. These samples are analysed 
to determine the \CP-violating parameters $x_\pm \equiv r_B \cos (\delta_B \pm \gamma)$ and $y_\pm \equiv r_B \sin (\delta_B \pm \gamma)$,
where $r_B$ is the ratio of the absolute values of the $B^+ \to D^0 K^-$ and $B^+ \to \Dzb K^-$ amplitudes,  $\delta_B$ is the strong-phase difference between them, and $\gamma$ is an angle of the unitarity triangle.   
The analysis is performed in bins of the $D$ decay Dalitz plot, and existing measurements of the CLEO-c experiment are used to
provide input on the $D$ decay strong-phase parameters $(c_i, s_i)$~\cite{CLEOCISI}.   Such an approach allows the analysis to be free from
any model-dependent assumptions on the strong-phase variation across the Dalitz plot.  
The following results are obtained:
\begin{align*}
x_+ &= (-7.7 \pm 2.4 \pm 1.0 \pm 0.4 )\times 10^{-2}, \;\,  &x_- &=  (2.5 \pm 2.5 \pm 1.0 \pm 0.5) \times 10^{-2}, \nonumber\\
y_+ &= (-2.2 \pm 2.5 \pm 0.4 \pm 1.0) \times 10^{-2}, \;\,  &y_- &=  (7.5 \pm 2.9 \pm 0.5 \pm 1.4) \times 10^{-2}, \nonumber
\end{align*}
where the first uncertainties are statistical, the second are systematic and the third arise from the experimental knowledge of the $(c_i, s_i)$ parameters. 
The results are the most precise values of these \CP observables obtained from a single measurement.

From the above results, the following values of the underlying physics parameters are derived:  $r_B = 0.080^{\,+0.019}_{\,-0.021}$, $\gamma = (62^{\,+15}_{\,-14})^\circ$ and $\delta_B = (134^{\,+14}_{\,-15})^\circ$.  
These values are consistent with the world averages of results from previous measurements~\cite{PDG2012}, but should not be combined with the model-dependent measurements~\cite{LHCb-PAPER-2014-017}. These values improve upon and supersede the results from a previous model-independent measurement performed with 1.0~${\rm fb^{-1}}$ of data collected by LHCb in 2011~\cite{LHCbGGSZ1fb}.

\section*{Acknowledgements}
 
\noindent We express our gratitude to our colleagues in the CERN
accelerator departments for the excellent performance of the LHC. We
thank the technical and administrative staff at the LHCb
institutes. We acknowledge support from CERN and from the national
agencies: CAPES, CNPq, FAPERJ and FINEP (Brazil); NSFC (China);
CNRS/IN2P3 (France); BMBF, DFG, HGF and MPG (Germany); SFI (Ireland); INFN (Italy); 
FOM and NWO (The Netherlands); MNiSW and NCN (Poland); MEN/IFA (Romania); 
MinES and FANO (Russia); MinECo (Spain); SNSF and SER (Switzerland); 
NASU (Ukraine); STFC (United Kingdom); NSF (USA).
The Tier1 computing centres are supported by IN2P3 (France), KIT and BMBF 
(Germany), INFN (Italy), NWO and SURF (The Netherlands), PIC (Spain), GridPP 
(United Kingdom).
We are indebted to the communities behind the multiple open 
source software packages on which we depend. We are also thankful for the 
computing resources and the access to software R\&D tools provided by Yandex LLC (Russia).
Individual groups or members have received support from 
EPLANET, Marie Sk\l{}odowska-Curie Actions and ERC (European Union), 
Conseil g\'{e}n\'{e}ral de Haute-Savoie, Labex ENIGMASS and OCEVU, 
R\'{e}gion Auvergne (France), RFBR (Russia), XuntaGal and GENCAT (Spain), Royal Society and Royal
Commission for the Exhibition of 1851 (United Kingdom).

\clearpage

\addcontentsline{toc}{section}{References}
\setboolean{inbibliography}{true}
\bibliographystyle{LHCb}
\bibliography{main,ana,LHCb-DP,LHCb-PAPER}

\ifx\mcitethebibliography\mciteundefinedmacro
\PackageError{LHCb.bst}{mciteplus.sty has not been loaded}
{This bibstyle requires the use of the mciteplus package.}\fi
\providecommand{\href}[2]{#2}
\begin{mcitethebibliography}{10}
\mciteSetBstSublistMode{n}
\mciteSetBstMaxWidthForm{subitem}{\alph{mcitesubitemcount})}
\mciteSetBstSublistLabelBeginEnd{\mcitemaxwidthsubitemform\space}
{\relax}{\relax}

\bibitem{LHCb-PAPER-2012-001}
LHCb collaboration, R.~Aaij {\em et~al.},
  \ifthenelse{\boolean{articletitles}}{{\it {Observation of $CP$ violation in
  $B^\pm \to D K^\pm$ decays}},
  }{}\href{http://dx.doi.org/10.1016/j.physletb.2012.04.060}{Phys.\ Lett.\
  {\bf B712} (2012) 203}, Erratum
  \href{http://dx.doi.org/10.1016/j.physletb.2012.05.060}{ibid.\   {\bf B713}
  (2012) 351}, \href{http://arxiv.org/abs/1203.3662}{{\tt
  arXiv:1203.3662}}\relax
\mciteBstWouldAddEndPuncttrue
\mciteSetBstMidEndSepPunct{\mcitedefaultmidpunct}
{\mcitedefaultendpunct}{\mcitedefaultseppunct}\relax
\EndOfBibitem
\bibitem{LHCb-PAPER-2012-055}
LHCb collaboration, R.~Aaij {\em et~al.},
  \ifthenelse{\boolean{articletitles}}{{\it {Observation of the suppressed ADS
  modes $B^\pm \to [\pi^\pm K^\mp\pi^+\pi^-]_D K^\pm$ and $B^\pm \to [\pi^\pm
  K^\mp \pi^+\pi^-]_D \pi^\pm$}},
  }{}\href{http://dx.doi.org/10.1016/j.physletb.2013.05.009}{Phys.\ Lett.\
  {\bf B723} (2013) 44}, \href{http://arxiv.org/abs/1303.4646}{{\tt
  arXiv:1303.4646}}\relax
\mciteBstWouldAddEndPuncttrue
\mciteSetBstMidEndSepPunct{\mcitedefaultmidpunct}
{\mcitedefaultendpunct}{\mcitedefaultseppunct}\relax
\EndOfBibitem
\bibitem{LHCb-PAPER-2013-068}
LHCb collaboration, R.~Aaij {\em et~al.},
  \ifthenelse{\boolean{articletitles}}{{\it {A study of CP violation in $B^\pm
  \to D K^\pm$ and $B^\pm \to D \pi^\pm$ decays with $D \to K^0_S K^\pm
  \pi^\mp$ final states}},
  }{}\href{http://dx.doi.org/10.1016/j.physletb.2014.03.051}{Phys.\ Lett.\
  {\bf B733} (2014) 36}, \href{http://arxiv.org/abs/1402.2982}{{\tt
  arXiv:1402.2982}}\relax
\mciteBstWouldAddEndPuncttrue
\mciteSetBstMidEndSepPunct{\mcitedefaultmidpunct}
{\mcitedefaultendpunct}{\mcitedefaultseppunct}\relax
\EndOfBibitem
\bibitem{LHCbGGSZ1fb}
LHCb collaboration, R.~Aaij {\em et~al.},
  \ifthenelse{\boolean{articletitles}}{{\it {A model-independent Dalitz plot
  analysis of $B^\pm \to D K^\pm$ with $D \to K^0_{\rm S} h^+h^-$ ($h=\pi, K$)
  decays and constraints on the CKM angle $\gamma$}},
  }{}\href{http://dx.doi.org/10.1016/j.physletb.2012.10.020}{Phys.\ Lett.\
  {\bf B718} (2012) 43}, \href{http://arxiv.org/abs/1209.5869}{{\tt
  arXiv:1209.5869}}\relax
\mciteBstWouldAddEndPuncttrue
\mciteSetBstMidEndSepPunct{\mcitedefaultmidpunct}
{\mcitedefaultendpunct}{\mcitedefaultseppunct}\relax
\EndOfBibitem
\bibitem{LHCb-PAPER-2014-028}
LHCb collaboration, R.~Aaij {\em et~al.},
  \ifthenelse{\boolean{articletitles}}{{\it {Measurements of CP violation
  parameters in $B^0\to DK^{*0}$ decays}},
  }{}\href{http://arxiv.org/abs/1407.8136}{{\tt arXiv:1407.8136}}, {submitted
  to Phys. Rev. D}\relax
\mciteBstWouldAddEndPuncttrue
\mciteSetBstMidEndSepPunct{\mcitedefaultmidpunct}
{\mcitedefaultendpunct}{\mcitedefaultseppunct}\relax
\EndOfBibitem
\bibitem{LHCb-PAPER-2014-038}
LHCb collaboration, R.~Aaij {\em et~al.},
  \ifthenelse{\boolean{articletitles}}{{\it {Measurement of CP asymmetry in
  $B^0_s \to D^\mp_s K^\pm$ decays}},
  }{}\href{http://arxiv.org/abs/1407.6127}{{\tt arXiv:1407.6127}}, {submitted
  to JHEP}\relax
\mciteBstWouldAddEndPuncttrue
\mciteSetBstMidEndSepPunct{\mcitedefaultmidpunct}
{\mcitedefaultendpunct}{\mcitedefaultseppunct}\relax
\EndOfBibitem
\bibitem{GGSZ}
A.~Giri, Y.~Grossman, A.~Soffer, and J.~Zupan,
  \ifthenelse{\boolean{articletitles}}{{\it {Determining $\gamma$ using
  ${B}^{\pm}\rightarrow DK^\pm$ with multibody $D$ decays}},
  }{}\href{http://dx.doi.org/10.1103/PhysRevD.68.054018}{Phys.\ Rev.\  {\bf
  D68} (2003) 054018}, \href{http://arxiv.org/abs/hep-ph/0303187}{{\tt
  arXiv:hep-ph/0303187}}\relax
\mciteBstWouldAddEndPuncttrue
\mciteSetBstMidEndSepPunct{\mcitedefaultmidpunct}
{\mcitedefaultendpunct}{\mcitedefaultseppunct}\relax
\EndOfBibitem
\bibitem{BONDARGGSZ}
A.~Bondar, {\em {Proceedings of BINP special analysis meeting on Dalitz
  analysis, 24-26 Sep. 2002, unpublished}}\relax
\mciteBstWouldAddEndPuncttrue
\mciteSetBstMidEndSepPunct{\mcitedefaultmidpunct}
{\mcitedefaultendpunct}{\mcitedefaultseppunct}\relax
\EndOfBibitem
\bibitem{BABAR2005}
\babar collaboration, B.~Aubert {\em et~al.},
  \ifthenelse{\boolean{articletitles}}{{\it {Measurement of the
  Cabibbo-Kobayashi-Maskawa angle $\gamma$ in $B^\mp \to D^{(*)} K^\mp$ decays
  with a Dalitz analysis of $D \to K^0_{\rm S} \pi^- \pi^+$}},
  }{}\href{http://dx.doi.org/10.1103/PhysRevLett.95.121802}{Phys.\ Rev.\ Lett.\
   {\bf 95} (2005) 121802}, \href{http://arxiv.org/abs/hep-ex/0504039}{{\tt
  arXiv:hep-ex/0504039}}\relax
\mciteBstWouldAddEndPuncttrue
\mciteSetBstMidEndSepPunct{\mcitedefaultmidpunct}
{\mcitedefaultendpunct}{\mcitedefaultseppunct}\relax
\EndOfBibitem
\bibitem{BABAR2008}
\babar collaboration, B.~Aubert {\em et~al.},
  \ifthenelse{\boolean{articletitles}}{{\it {Improved measurement of the CKM
  angle $\gamma$ in $B^\mp \to D^{(*)} K^{(*\mp)}$ decays with a Dalitz plot
  analysis of $D$ decays to $K^0_{\rm S} \pi^{+} \pi^{-}$ and $K^0_{\rm S}
  K^{+} K^{-}$}}, }{}\href{http://dx.doi.org/10.1103/PhysRevD.78.034023}{Phys.\
  Rev.\  {\bf D78} (2008) 034023}, \href{http://arxiv.org/abs/0804.2089}{{\tt
  arXiv:0804.2089}}\relax
\mciteBstWouldAddEndPuncttrue
\mciteSetBstMidEndSepPunct{\mcitedefaultmidpunct}
{\mcitedefaultendpunct}{\mcitedefaultseppunct}\relax
\EndOfBibitem
\bibitem{BABAR2010}
\babar collaboration, P.~del Amo~Sanchez {\em et~al.},
  \ifthenelse{\boolean{articletitles}}{{\it {Evidence for direct CP violation
  in the measurement of the Cabibbo-Kobayashi-Maskawa angle $\gamma$ with
  $B^\mp \to D^{(*)} K^{(*)\mp}$ decays}},
  }{}\href{http://dx.doi.org/10.1103/PhysRevLett.105.121801}{Phys.\ Rev.\
  Lett.\  {\bf 105} (2010) 121801}, \href{http://arxiv.org/abs/1005.1096}{{\tt
  arXiv:1005.1096}}\relax
\mciteBstWouldAddEndPuncttrue
\mciteSetBstMidEndSepPunct{\mcitedefaultmidpunct}
{\mcitedefaultendpunct}{\mcitedefaultseppunct}\relax
\EndOfBibitem
\bibitem{BELLE2004}
Belle collaboration, A.~Poluektov {\em et~al.},
  \ifthenelse{\boolean{articletitles}}{{\it {Measurement of $\phi_3$ with
  Dalitz plot analysis of $B^\pm \to D^{(*)} K^\pm$ decay}},
  }{}\href{http://dx.doi.org/10.1103/PhysRevD.70.072003}{Phys.\ Rev.\  {\bf
  D70} (2004) 072003}, \href{http://arxiv.org/abs/hep-ex/0406067}{{\tt
  arXiv:hep-ex/0406067}}\relax
\mciteBstWouldAddEndPuncttrue
\mciteSetBstMidEndSepPunct{\mcitedefaultmidpunct}
{\mcitedefaultendpunct}{\mcitedefaultseppunct}\relax
\EndOfBibitem
\bibitem{BELLE2006}
Belle collaboration, A.~Poluektov {\em et~al.},
  \ifthenelse{\boolean{articletitles}}{{\it {Measurement of $\phi_3$ with
  Dalitz plot analysis of $B^+ \to D^{(*)}K^{(*)+}$ decay}},
  }{}\href{http://dx.doi.org/10.1103/PhysRevD.73.112009}{Phys.\ Rev.\  {\bf
  D73} (2006) 112009}, \href{http://arxiv.org/abs/hep-ex/0604054}{{\tt
  arXiv:hep-ex/0604054}}\relax
\mciteBstWouldAddEndPuncttrue
\mciteSetBstMidEndSepPunct{\mcitedefaultmidpunct}
{\mcitedefaultendpunct}{\mcitedefaultseppunct}\relax
\EndOfBibitem
\bibitem{BELLE2010}
Belle collaboration, A.~Poluektov {\em et~al.},
  \ifthenelse{\boolean{articletitles}}{{\it {Evidence for direct CP violation
  in the decay $B \to D^{(*)}K$, $D \to K^0_{\rm S} \pi^+ \pi^-$ and
  measurement of the CKM phase $\phi_3$}},
  }{}\href{http://dx.doi.org/10.1103/PhysRevD.81.112002}{Phys.\ Rev.\  {\bf
  D81} (2010) 112002}, \href{http://arxiv.org/abs/1003.3360}{{\tt
  arXiv:1003.3360}}\relax
\mciteBstWouldAddEndPuncttrue
\mciteSetBstMidEndSepPunct{\mcitedefaultmidpunct}
{\mcitedefaultendpunct}{\mcitedefaultseppunct}\relax
\EndOfBibitem
\bibitem{LHCb-PAPER-2014-017}
LHCb collaboration, R.~Aaij {\em et~al.},
  \ifthenelse{\boolean{articletitles}}{{\it {Measurement of $CP$ violation and
  constraints on the CKM angle $\gamma$ in $B^\pm \to D K^\pm$ with $D \to
  K_S^0\pi^+\pi^-$ decays}}, }{}\href{http://arxiv.org/abs/1407.6211}{{\tt
  arXiv:1407.6211}}, {submitted to Nucl. Phys. B}\relax
\mciteBstWouldAddEndPuncttrue
\mciteSetBstMidEndSepPunct{\mcitedefaultmidpunct}
{\mcitedefaultendpunct}{\mcitedefaultseppunct}\relax
\EndOfBibitem
\bibitem{BPMODIND1}
A.~Bondar and A.~Poluektov, \ifthenelse{\boolean{articletitles}}{{\it
  {Feasibility study of model-independent approach to $\phi_3$ measurement
  using Dalitz plot analysis}},
  }{}\href{http://dx.doi.org/10.1140/epjc/s2006-02590-x}{Eur.\ Phys.\ J.\  {\bf
  C47} (2006) 347}, \href{http://arxiv.org/abs/hep-ph/0510246}{{\tt
  arXiv:hep-ph/0510246}}\relax
\mciteBstWouldAddEndPuncttrue
\mciteSetBstMidEndSepPunct{\mcitedefaultmidpunct}
{\mcitedefaultendpunct}{\mcitedefaultseppunct}\relax
\EndOfBibitem
\bibitem{BPMODIND2}
A.~Bondar and A.~Poluektov, \ifthenelse{\boolean{articletitles}}{{\it {The use
  of quantum-correlated $D^0$ decays for $\phi_3$ measurement}},
  }{}\href{http://dx.doi.org/10.1140/epjc/s10052-008-0600-z}{Eur.\ Phys.\ J.\
  {\bf C55} (2008) 51}, \href{http://arxiv.org/abs/0801.0840}{{\tt
  arXiv:0801.0840}}\relax
\mciteBstWouldAddEndPuncttrue
\mciteSetBstMidEndSepPunct{\mcitedefaultmidpunct}
{\mcitedefaultendpunct}{\mcitedefaultseppunct}\relax
\EndOfBibitem
\bibitem{CLEOCISI}
CLEO collaboration, J.~Libby {\em et~al.},
  \ifthenelse{\boolean{articletitles}}{{\it {Model-independent determination of
  the strong-phase difference between $D^0$ and $\overline{D}^0 \to
  K^0_{\rm{S,L}} h^+ h^-$ ($h=\pi,K$) and its impact on the measurement of the
  CKM angle $\gamma/\phi_3$}},
  }{}\href{http://dx.doi.org/10.1103/PhysRevD.82.112006}{Phys.\ Rev.\  {\bf
  D82} (2010) 112006}, \href{http://arxiv.org/abs/1010.2817}{{\tt
  arXiv:1010.2817}}\relax
\mciteBstWouldAddEndPuncttrue
\mciteSetBstMidEndSepPunct{\mcitedefaultmidpunct}
{\mcitedefaultendpunct}{\mcitedefaultseppunct}\relax
\EndOfBibitem
\bibitem{BELLEMODIND}
Belle collaboration, H.~Aihara {\em et~al.},
  \ifthenelse{\boolean{articletitles}}{{\it {First measurement of $\phi_3$ with
  a model-independent Dalitz plot analysis of $B^\pm \to DK^\pm$ , $D \to
  K^0_{\rm S} \pi^+ \pi^-$ decay}},
  }{}\href{http://dx.doi.org/10.1103/PhysRevD.85.112014}{Phys.\ Rev.\  {\bf
  D85} (2012) 112014}, \href{http://arxiv.org/abs/1204.6561}{{\tt
  arXiv:1204.6561}}\relax
\mciteBstWouldAddEndPuncttrue
\mciteSetBstMidEndSepPunct{\mcitedefaultmidpunct}
{\mcitedefaultendpunct}{\mcitedefaultseppunct}\relax
\EndOfBibitem
\bibitem{PDG2012}
Particle Data Group, J.~Beringer {\em et~al.},
  \ifthenelse{\boolean{articletitles}}{{\it {\href{http://pdg.lbl.gov/}{Review
  of particle physics}}},
  }{}\href{http://dx.doi.org/10.1103/PhysRevD.86.010001}{Phys.\ Rev.\  {\bf
  D86} (2012) 010001}, {and 2013 partial update for the 2014 edition}\relax
\mciteBstWouldAddEndPuncttrue
\mciteSetBstMidEndSepPunct{\mcitedefaultmidpunct}
{\mcitedefaultendpunct}{\mcitedefaultseppunct}\relax
\EndOfBibitem
\bibitem{BPV}
A.~Bondar, A.~Poluektov, and V.~Vorobiev,
  \ifthenelse{\boolean{articletitles}}{{\it {Charm mixing in a
  model-independent analysis of correlated $D^0 \overline{D}^0$ decays}},
  }{}\href{http://dx.doi.org/10.1103/PhysRevD.82.034033}{Phys.\ Rev.\  {\bf
  D82} (2010) 034033}, \href{http://arxiv.org/abs/1004.2350}{{\tt
  arXiv:1004.2350}}\relax
\mciteBstWouldAddEndPuncttrue
\mciteSetBstMidEndSepPunct{\mcitedefaultmidpunct}
{\mcitedefaultendpunct}{\mcitedefaultseppunct}\relax
\EndOfBibitem
\bibitem{Yuval}
Y.~Grossman and M.~Savastio, \ifthenelse{\boolean{articletitles}}{{\it {Effects
  of \Kz--\Kzb mixing on determining $\gamma$ from $B^\pm \to DK^\pm$}},
  }{}\href{http://dx.doi.org/10.1007/JHEP03(2014)008}{JHEP {\bf 03} (2014)
  008}, \href{http://arxiv.org/abs/1311.3575}{{\tt arXiv:1311.3575}}\relax
\mciteBstWouldAddEndPuncttrue
\mciteSetBstMidEndSepPunct{\mcitedefaultmidpunct}
{\mcitedefaultendpunct}{\mcitedefaultseppunct}\relax
\EndOfBibitem
\bibitem{LHCb-PAPER-2014-013}
LHCb collaboration, R.~Aaij {\em et~al.},
  \ifthenelse{\boolean{articletitles}}{{\it {Measurement of $CP$ asymmetry in
  $D^0 \to K^- K^+$ and $D^0 \to \pi^- \pi^+$ decays}},
  }{}\href{http://dx.doi.org/10.1007/JHEP07(2014)041}{JHEP {\bf 07} (2014)
  041}, \href{http://arxiv.org/abs/1405.2797}{{\tt arXiv:1405.2797}}\relax
\mciteBstWouldAddEndPuncttrue
\mciteSetBstMidEndSepPunct{\mcitedefaultmidpunct}
{\mcitedefaultendpunct}{\mcitedefaultseppunct}\relax
\EndOfBibitem
\bibitem{Alves:2008zz}
LHCb collaboration, A.~A. Alves~Jr.\ {\em et~al.},
  \ifthenelse{\boolean{articletitles}}{{\it {The \lhcb detector at the LHC}},
  }{}\href{http://dx.doi.org/10.1088/1748-0221/3/08/S08005}{JINST {\bf 3}
  (2008) S08005}\relax
\mciteBstWouldAddEndPuncttrue
\mciteSetBstMidEndSepPunct{\mcitedefaultmidpunct}
{\mcitedefaultendpunct}{\mcitedefaultseppunct}\relax
\EndOfBibitem
\bibitem{LHCb-DP-2013-003}
R.~Arink {\em et~al.}, \ifthenelse{\boolean{articletitles}}{{\it {Performance
  of the LHCb Outer Tracker}},
  }{}\href{http://dx.doi.org/10.1088/1748-0221/9/01/P01002}{JINST {\bf 9}
  (2014) P01002}, \href{http://arxiv.org/abs/1311.3893}{{\tt
  arXiv:1311.3893}}\relax
\mciteBstWouldAddEndPuncttrue
\mciteSetBstMidEndSepPunct{\mcitedefaultmidpunct}
{\mcitedefaultendpunct}{\mcitedefaultseppunct}\relax
\EndOfBibitem
\bibitem{LHCb-DP-2012-003}
M.~Adinolfi {\em et~al.}, \ifthenelse{\boolean{articletitles}}{{\it
  {Performance of the \lhcb RICH detector at the LHC}},
  }{}\href{http://dx.doi.org/10.1140/epjc/s10052-013-2431-9}{Eur.\ Phys.\ J.\
  {\bf C73} (2013) 2431}, \href{http://arxiv.org/abs/1211.6759}{{\tt
  arXiv:1211.6759}}\relax
\mciteBstWouldAddEndPuncttrue
\mciteSetBstMidEndSepPunct{\mcitedefaultmidpunct}
{\mcitedefaultendpunct}{\mcitedefaultseppunct}\relax
\EndOfBibitem
\bibitem{LHCb-DP-2012-002}
A.~A. Alves~Jr.\ {\em et~al.}, \ifthenelse{\boolean{articletitles}}{{\it
  {Performance of the LHCb muon system}},
  }{}\href{http://dx.doi.org/10.1088/1748-0221/8/02/P02022}{JINST {\bf 8}
  (2013) P02022}, \href{http://arxiv.org/abs/1211.1346}{{\tt
  arXiv:1211.1346}}\relax
\mciteBstWouldAddEndPuncttrue
\mciteSetBstMidEndSepPunct{\mcitedefaultmidpunct}
{\mcitedefaultendpunct}{\mcitedefaultseppunct}\relax
\EndOfBibitem
\bibitem{LHCb-DP-2012-004}
R.~Aaij {\em et~al.}, \ifthenelse{\boolean{articletitles}}{{\it {The \lhcb
  trigger and its performance in 2011}},
  }{}\href{http://dx.doi.org/10.1088/1748-0221/8/04/P04022}{JINST {\bf 8}
  (2013) P04022}, \href{http://arxiv.org/abs/1211.3055}{{\tt
  arXiv:1211.3055}}\relax
\mciteBstWouldAddEndPuncttrue
\mciteSetBstMidEndSepPunct{\mcitedefaultmidpunct}
{\mcitedefaultendpunct}{\mcitedefaultseppunct}\relax
\EndOfBibitem
\bibitem{Sjostrand:2006za}
T.~Sj\"{o}strand, S.~Mrenna, and P.~Skands,
  \ifthenelse{\boolean{articletitles}}{{\it {PYTHIA 6.4 physics and manual}},
  }{}\href{http://dx.doi.org/10.1088/1126-6708/2006/05/026}{JHEP {\bf 05}
  (2006) 026}, \href{http://arxiv.org/abs/hep-ph/0603175}{{\tt
  arXiv:hep-ph/0603175}}\relax
\mciteBstWouldAddEndPuncttrue
\mciteSetBstMidEndSepPunct{\mcitedefaultmidpunct}
{\mcitedefaultendpunct}{\mcitedefaultseppunct}\relax
\EndOfBibitem
\bibitem{Sjostrand:2007gs}
T.~Sj\"{o}strand, S.~Mrenna, and P.~Skands,
  \ifthenelse{\boolean{articletitles}}{{\it {A brief introduction to PYTHIA
  8.1}}, }{}\href{http://dx.doi.org/10.1016/j.cpc.2008.01.036}{Comput.\ Phys.\
  Commun.\  {\bf 178} (2008) 852}, \href{http://arxiv.org/abs/0710.3820}{{\tt
  arXiv:0710.3820}}\relax
\mciteBstWouldAddEndPuncttrue
\mciteSetBstMidEndSepPunct{\mcitedefaultmidpunct}
{\mcitedefaultendpunct}{\mcitedefaultseppunct}\relax
\EndOfBibitem
\bibitem{LHCb-PROC-2010-056}
I.~Belyaev {\em et~al.}, \ifthenelse{\boolean{articletitles}}{{\it {Handling of
  the generation of primary events in \gauss, the \lhcb simulation framework}},
  }{}\href{http://dx.doi.org/10.1109/NSSMIC.2010.5873949}{Nuclear Science
  Symposium Conference Record (NSS/MIC) {\bf IEEE} (2010) 1155}\relax
\mciteBstWouldAddEndPuncttrue
\mciteSetBstMidEndSepPunct{\mcitedefaultmidpunct}
{\mcitedefaultendpunct}{\mcitedefaultseppunct}\relax
\EndOfBibitem
\bibitem{Lange:2001uf}
D.~J. Lange, \ifthenelse{\boolean{articletitles}}{{\it {The EvtGen particle
  decay simulation package}},
  }{}\href{http://dx.doi.org/10.1016/S0168-9002(01)00089-4}{Nucl.\ Instrum.\
  Meth.\  {\bf A462} (2001) 152}\relax
\mciteBstWouldAddEndPuncttrue
\mciteSetBstMidEndSepPunct{\mcitedefaultmidpunct}
{\mcitedefaultendpunct}{\mcitedefaultseppunct}\relax
\EndOfBibitem
\bibitem{Golonka:2005pn}
P.~Golonka and Z.~Was, \ifthenelse{\boolean{articletitles}}{{\it {PHOTOS Monte
  Carlo: a precision tool for QED corrections in $Z$ and $W$ decays}},
  }{}\href{http://dx.doi.org/10.1140/epjc/s2005-02396-4}{Eur.\ Phys.\ J.\  {\bf
  C45} (2006) 97}, \href{http://arxiv.org/abs/hep-ph/0506026}{{\tt
  arXiv:hep-ph/0506026}}\relax
\mciteBstWouldAddEndPuncttrue
\mciteSetBstMidEndSepPunct{\mcitedefaultmidpunct}
{\mcitedefaultendpunct}{\mcitedefaultseppunct}\relax
\EndOfBibitem
\bibitem{Allison:2006ve}
Geant4 collaboration, J.~Allison {\em et~al.},
  \ifthenelse{\boolean{articletitles}}{{\it {Geant4 developments and
  applications}}, }{}\href{http://dx.doi.org/10.1109/TNS.2006.869826}{IEEE
  Trans.\ Nucl.\ Sci.\  {\bf 53} (2006) 270}\relax
\mciteBstWouldAddEndPuncttrue
\mciteSetBstMidEndSepPunct{\mcitedefaultmidpunct}
{\mcitedefaultendpunct}{\mcitedefaultseppunct}\relax
\EndOfBibitem
\bibitem{Agostinelli:2002hh}
Geant4 collaboration, S.~Agostinelli {\em et~al.},
  \ifthenelse{\boolean{articletitles}}{{\it {Geant4: a simulation toolkit}},
  }{}\href{http://dx.doi.org/10.1016/S0168-9002(03)01368-8}{Nucl.\ Instrum.\
  Meth.\  {\bf A506} (2003) 250}\relax
\mciteBstWouldAddEndPuncttrue
\mciteSetBstMidEndSepPunct{\mcitedefaultmidpunct}
{\mcitedefaultendpunct}{\mcitedefaultseppunct}\relax
\EndOfBibitem
\bibitem{LHCb-PROC-2011-006}
M.~Clemencic {\em et~al.}, \ifthenelse{\boolean{articletitles}}{{\it {The \lhcb
  simulation application, \gauss: design, evolution and experience}},
  }{}\href{http://dx.doi.org/10.1088/1742-6596/331/3/032023}{{J.\ Phys.\ Conf.\
  Ser.\ } {\bf 331} (2011) 032023}\relax
\mciteBstWouldAddEndPuncttrue
\mciteSetBstMidEndSepPunct{\mcitedefaultmidpunct}
{\mcitedefaultendpunct}{\mcitedefaultseppunct}\relax
\EndOfBibitem
\bibitem{BBDT}
V.~V. Gligorov and M.~Williams, \ifthenelse{\boolean{articletitles}}{{\it
  {Efficient, reliable and fast high-level triggering using a bonsai boosted
  decision tree}},
  }{}\href{http://dx.doi.org/10.1088/1748-0221/8/02/P02013}{JINST {\bf 8}
  (2013) P02013}, \href{http://arxiv.org/abs/1210.6861}{{\tt
  arXiv:1210.6861}}\relax
\mciteBstWouldAddEndPuncttrue
\mciteSetBstMidEndSepPunct{\mcitedefaultmidpunct}
{\mcitedefaultendpunct}{\mcitedefaultseppunct}\relax
\EndOfBibitem
\bibitem{Breiman}
L.~Breiman, J.~H. Friedman, R.~A. Olshen, and C.~J. Stone, {\em Classification
  and regression trees}, Wadsworth international group, Belmont, California,
  USA, 1984\relax
\mciteBstWouldAddEndPuncttrue
\mciteSetBstMidEndSepPunct{\mcitedefaultmidpunct}
{\mcitedefaultendpunct}{\mcitedefaultseppunct}\relax
\EndOfBibitem
\bibitem{AdaBoost}
R.~E. Schapire and Y.~Freund, \ifthenelse{\boolean{articletitles}}{{\it A
  decision-theoretic generalization of on-line learning and an application to
  boosting}, }{}\href{http://dx.doi.org/10.1006/jcss.1997.1504}{Jour.\ Comp.\
  and Syst.\ Sc.\  {\bf 55} (1997) 119}\relax
\mciteBstWouldAddEndPuncttrue
\mciteSetBstMidEndSepPunct{\mcitedefaultmidpunct}
{\mcitedefaultendpunct}{\mcitedefaultseppunct}\relax
\EndOfBibitem
\bibitem{Hulsbergen:2005pu}
W.~D. Hulsbergen, \ifthenelse{\boolean{articletitles}}{{\it {Decay chain
  fitting with a Kalman filter}},
  }{}\href{http://dx.doi.org/10.1016/j.nima.2005.06.078}{Nucl.\ Instrum.\
  Meth.\  {\bf A552} (2005) 566},
  \href{http://arxiv.org/abs/physics/0503191}{{\tt
  arXiv:physics/0503191}}\relax
\mciteBstWouldAddEndPuncttrue
\mciteSetBstMidEndSepPunct{\mcitedefaultmidpunct}
{\mcitedefaultendpunct}{\mcitedefaultseppunct}\relax
\EndOfBibitem
\bibitem{Pivk:2004ty}
M.~Pivk and F.~R. Le~Diberder, \ifthenelse{\boolean{articletitles}}{{\it
  {sPlot: a statistical tool to unfold data distributions}},
  }{}\href{http://dx.doi.org/10.1016/j.nima.2005.08.106}{Nucl.\ Instrum.\
  Meth.\  {\bf A555} (2005) 356},
  \href{http://arxiv.org/abs/physics/0402083}{{\tt
  arXiv:physics/0402083}}\relax
\mciteBstWouldAddEndPuncttrue
\mciteSetBstMidEndSepPunct{\mcitedefaultmidpunct}
{\mcitedefaultendpunct}{\mcitedefaultseppunct}\relax
\EndOfBibitem
\bibitem{HFAG}
Heavy Flavor Averaging Group, D.~Asner {\em et~al.},
  \ifthenelse{\boolean{articletitles}}{{\it {Averages of b-hadron, c-hadron,
  and $\tau$-lepton properties}}, }{}\href{http://arxiv.org/abs/1010.1589}{{\tt
  arXiv:1010.1589}}, Updates available online at
  \href{http://www.slac.stanford.edu/xorg/hfag}{\url{http://www.slac.stanford.%
edu/xorg/hfag}}\relax
\mciteBstWouldAddEndPuncttrue
\mciteSetBstMidEndSepPunct{\mcitedefaultmidpunct}
{\mcitedefaultendpunct}{\mcitedefaultseppunct}\relax
\EndOfBibitem
\bibitem{FELDMANCOUSINS}
G.~J. Feldman and R.~D. Cousins, \ifthenelse{\boolean{articletitles}}{{\it {A
  unified approach to the classical statistical analysis of small signals}},
  }{}\href{http://dx.doi.org/10.1103/PhysRevD.57.3873}{Phys.\ Rev.\  {\bf D57}
  (1998) 3873}, \href{http://arxiv.org/abs/physics/9711021}{{\tt
  arXiv:physics/9711021}}\relax
\mciteBstWouldAddEndPuncttrue
\mciteSetBstMidEndSepPunct{\mcitedefaultmidpunct}
{\mcitedefaultendpunct}{\mcitedefaultseppunct}\relax
\EndOfBibitem
\end{mcitethebibliography}

\newpage
\centerline{\large\bf LHCb collaboration}
\begin{flushleft}
\small
R.~Aaij$^{41}$, 
B.~Adeva$^{37}$, 
M.~Adinolfi$^{46}$, 
A.~Affolder$^{52}$, 
Z.~Ajaltouni$^{5}$, 
S.~Akar$^{6}$, 
J.~Albrecht$^{9}$, 
F.~Alessio$^{38}$, 
M.~Alexander$^{51}$, 
S.~Ali$^{41}$
% ...

{\footnotesize \it
% ...
$ ^{5}$Clermont Universit\'{e}, Universit\'{e} Blaise Pascal, CNRS/IN2P3, LPC, Clermont-Ferrand, France\\
$ ^{6}$CPPM, Aix-Marseille Universit\'{e}, CNRS/IN2P3, Marseille, France\\
% ...
$ ^{9}$Fakult\"{a}t Physik, Technische Universit\"{a}t Dortmund, Dortmund, Germany\\
% ...
$ ^{37}$Universidad de Santiago de Compostela, Santiago de Compostela, Spain\\
$ ^{38}$European Organization for Nuclear Research (CERN), Geneva, Switzerland\\
% ...
$ ^{41}$Nikhef National Institute for Subatomic Physics, Amsterdam, The Netherlands\\
% ...
$ ^{46}$H.H. Wills Physics Laboratory, University of Bristol, Bristol, United Kingdom\\
% ...
$ ^{51}$School of Physics and Astronomy, University of Glasgow, Glasgow, United Kingdom\\
$ ^{52}$Oliver Lodge Laboratory, University of Liverpool, Liverpool, United Kingdom\\
% ...
}
\end{flushleft}

\end{document}